\documentclass{sig-alternate-2013}

\newfont{\mycrnotice}{ptmr8t at 7pt}
\newfont{\myconfname}{ptmri8t at 7pt}
\let\crnotice\mycrnotice%
\let\confname\myconfname%

\makeatletter
\let\@copyrightspace\relax
\makeatother

\usepackage{microtype} %
\usepackage{graphicx}
\usepackage{times}
\usepackage{subfigure}

\usepackage{xspace}
\usepackage[bf]{caption}

\usepackage{booktabs}

\usepackage{paralist}
\usepackage{enumitem}

\usepackage{subfigure}
\usepackage{multirow}
\usepackage[hang,flushmargin]{footmisc}
\usepackage{algpseudocode}
\usepackage[hyphens]{url}
\usepackage{balance}
\usepackage{color}

\renewcommand{\footnotesize}{\fontsize{9}{10}\selectfont}

\makeatletter
\def\url@leostyle{%
  \@ifundefined{selectfont}{\def\UrlFont{}}%
  {\def\UrlFont{}}%
}
\makeatother	
\usepackage{breakurl}
\usepackage{fancyvrb}
\usepackage{amssymb}

\newcommand{\red}[1] {{\color{black}{#1}}}

\newcommand{\descr}[1]{\smallskip\noindent{\bf #1}}

\def\df{$D_{full}$\xspace}
\def\dd{$D_{denied}$\xspace}

\def\du{$D_{user}$\xspace}
\def\ds{$D_{sample}$\xspace}
\def\dipv4{$D_{IPv4}$\xspace}

\usepackage{listings}
\newcommand{\policydenied}{\emph{policy\_denied}\xspace}
\newcommand{\policyredirect}{\emph{policy\_redirect}\xspace}

\usepackage{paralist}

\begin{document}

\title{Censorship in the Wild:\hspace{-0.1cm} Analyzing Internet Filtering in Syria}

\numberofauthors{6} 
\author{
\alignauthor{
Abdelberi Chaabane\\
       \affaddr{INRIA Rh\^{o}ne-Alpes}\\
       \affaddr{Montbonnot, France}}
\alignauthor{
Terence Chen\\
       \affaddr{NICTA}\\
       \affaddr{Sydney, Australia}}
\alignauthor{
Mathieu Cunche\\
       \affaddr{University of Lyon \& INRIA}\\
       \affaddr{Lyon, France}}
\and
\alignauthor{
Emiliano De Cristofaro\\
       \affaddr{University College London}\\
       \affaddr{London, United Kingdom}}
\alignauthor{
Arik Friedman\\
       \affaddr{NICTA}\\
       \affaddr{Sydney, Australia}}
\alignauthor{
Mohamed Ali Kaafar\\
       \affaddr{NICTA \& INRIA Rh\^{o}ne-Alpes}\\
       \affaddr{Sydney, Australia}}
}

\maketitle

\begin{abstract}
Internet censorship is enforced by numerous governments worldwide, however, due to the lack of publicly available information, as well as the inherent risks of performing active measurements, it is often hard for the research community to investigate censorship practices in the wild. Thus, the leak of 600GB worth of logs from 7 Blue Coat SG-9000 proxies, deployed in Syria to filter Internet traffic at a country scale, represents a unique opportunity to provide a detailed snapshot of a real-world censorship ecosystem. 

This paper presents the methodology and the results of a measurement analysis of the leaked Blue Coat logs, revealing a relatively stealthy, yet quite targeted, censorship. We find that traffic is filtered in several ways: using IP addresses and domain names to block subnets or websites, and keywords or categories to target specific content. We show that keyword-based censorship produces some collateral damage as many requests are blocked even if they do not relate to sensitive content.  We also discover that Instant Messaging is heavily censored, while filtering of social media is limited to specific pages. Finally, we show that Syrian users try to evade censorship by using web/socks proxies, Tor, VPNs, and BitTorrent. To the best of our knowledge, our work provides the first analytical look into Internet filtering in Syria.
\end{abstract}

\section{Introduction}
As the relation between society and technology evolves, so does censorship---the practice of suppressing ideas and information that certain individuals, groups or government officials may find objectionable, dangerous, or detrimental.
Censors increasingly target access to, and dissemination of, electronic information, for instance, 
aiming to restrict freedom of speech, control knowledge available to the masses, or enforce religious/ethical principles. 

Even though the research community dedicated a lot of attention to censorship and its circumvention, knowledge and understanding of filtering technologies is inherently limited, as it is challenging and risky to conduct measurements from countries operating censorship, while logs pertaining to filtered traffic are obviously hard to come by.
Prior work has analyzed censorship practices in China~\cite{king2012censorship,knockel2011three,park2010empirical,winter2012great,xu2011internet}, Iran~\cite{anderson2012hidden,aryan2013internet,verkamp2012inferring}, 
Pakistan~\cite{nabi2013anatomy}, and a few Arab countries~\cite{dalek2013method},
however, mostly based on probing, i.e., {\em inferring} what information is being censored by generating requests and observing what content is blocked.  While providing valuable insights, these methods suffer from two main limitations: (1) only a limited number of requests can be observed, thus providing a skewed representation of the censorship policies due to the inability to enumerate all censored keywords, and (2) it is hard to assess the actual extent of the censorship, e.g., 
what kind/proportion of the overall traffic is being censored.

\descr{Roadmap.} In this paper, we present a measurement analysis of Internet filtering in Syria: we study a set of logs extracted from 7 Blue Coat SG-9000 proxies, which were deployed to monitor, filter and block traffic of Syrian users. The logs (600GB worth of data) were leaked by a ``hacktivist'' group, Telecomix, in October 2011, and relate to a period of 9 days between July and August 2011~\cite{leak}.
By analyzing the logs, we provide a detailed snapshot of how censorship was operated in Syria. %
As opposed to probing-based methods, the analysis of actual logs allows us to extract information about processed requests for \textit{both} censored and allowed traffic and provide a detailed snapshot of Syrian censorship practices. 

\descr{Main Findings.}  Our measurement-based analysis uncovers several interesting findings. First, we observe that a few different techniques are employed by Syrian authorities: IP-based filtering to block access to entire subnets (e.g., Israel), domain-based to block specific websites, keyword-based to target specific kinds of traffic (e.g., filtering-evading technologies, such as socks proxies), and category-based to target specific content and pages.
As a side-effect of keyword-based censorship, many requests are blocked even if they do not relate to any sensitive content or anti-censorship technologies. Such collateral damage affects Google toolbar's queries as well as a few ads delivery networks that are blocked as they generate requests containing the word {\em proxy}.

Also, the logs highlight that Instant Messaging software (like Skype) is heavily censored, while filtering of social media is limited to specific pages. Actually, most social networks (e.g., Facebook and Twitter) are not blocked, but only certain pages/groups are (e.g., the Syrian Revolution Facebook page).
We find that proxies have specialized roles and/or slightly different configurations, as some of them tend to censor more traffic than others. For instance, one particular proxy blocks Tor traffic for several days, while other proxies do not.
Finally, we show that Syrian Internet users not only try to evade censorship and surveillance using well-known web/socks proxies, Tor, and VPN software, but also use P2P file sharing software (BitTorrent) to fetch censored content. %

Our analysis shows that, compared to other countries (e.g., China and Iran), censorship in Syria seems to be less pervasive, yet quite targeted. Syrian censors particularly target Instant Messaging, information related to the political opposition (e.g., pages related to the ``Syrian Revolution''), and Israeli subnets.
Arguably, less evident censorship does not necessarily mean minor information control or less ubiquitous surveillance. In fact, Syrian users seem to be aware of this and do resort to censorship- and surveillance-evading software, as we show later in the paper, which seems to confirm reports about Syrian users enganging in self-censorship to avoid being arrested~\cite{selfcensorship,bbc,oni}.  

\descr{Contributions.} To the best of our knowledge, we provide the first detailed snapshot of Internet filtering in Syria and the first set of large-scale measurements of actual filtering proxies' logs.
We show how censorship was operated via a statistical overview of censorship activities and an analysis of temporal patterns, proxy specializations, and filtering of social network sites. Finally, we provide some details on the usage of surveillance- and  censorship-evading tools.

\descr{Remarks.} Logs studied in this paper date back to July-August 2011, thus, our work is not intended to provide insights to the {\em current} situation in Syria as censorship might have evolved in the last two years. According to Bloomberg~\cite{bloomberg}, Syria has invested \$500K in surveillance equipment in late 2011, thus an even more powerful filtering architecture might now be in place. Starting December 2012, Tor relays and bridges have reportedly been blocked~\cite{tor-syria}.
Nonetheless, our analysis uncovers methods that might still be in place, e.g., based on Deep Packet Inspection. Also, Blue Coat proxies are reportedly still used in several countries~\cite{citizenlab}. 

Our work serves as a case study of censorship in practice: it provides a first-of-its-kind, data-driven analysis of a real-world censorship ecosystem, exposing its underlying techniques, as well as its strengths and weaknesses, which we hope will facilitate the design of censorship-evading tools.

\descr{Paper Organization.} The rest of this paper is organized as follows. The next section reviews related work. Then, Section~\ref{sec:background} provides some background information and introduces the datasets studied throughout the paper.
Section~\ref{sec:general_characteristics} presents a statistical overview of Internet censorship in Syria based on the Blue Coat logs, while Section~\ref{sec:understanding_censorhip} provides a thorough analysis to better understand censorship practices. After focusing on social network sites in Section~\ref{sec:social_media} and anti-censorship technologies in Section~\ref{sec:anti}, we discuss our findings in Section~\ref{sec:discussion}. 
The paper concludes with Section~\ref{sec:conclusion}.

\section{Related Work}\label{sec:related}
Due to the limited availability of publicly available data, there is little prior work analyzing logs from filtering devices. A fairly large body of work, which we overview in this section, has focused on understanding and characterizing censorship processes via probing. By contrast, our work is really the first to analyze traffic observed by actual filtering proxies and to provide a detailed measurement-based snapshot of Syria's censorship infrastructure.

Ayran et al.~\cite{aryan2013internet} present measurements from an Iranian ISP, analyzing HTTP host-based blocking, keyword filtering, DNS hijacking, and protocol-based throttling, and conclude that the censorship infrastructure heavily relies on centralized equipment. 
Winter and Lindskog~\cite{winter2012great} conduct 
some measurements on traffic routed through Tor bridges/relays to understand how China blocks Tor. %
Also, Dainotti et al.~\cite{dainotti2011analysis} analyze country-wide Internet outages, in Egypt and Libya, using publicly available data such as BGP inter-domain routing control plane data.

Another line of work deals with fingerprinting and inferring censorship methods and equipments. Researchers from the Citizen Lab~\cite{citizenlab}
focus on censorship/surveillance performed using Blue Coat devices
and uncover 61 Blue Coat ProxySG devices and 316 Blue Coat PacketShaper appliances in 
24 different countries.
Dalek et al.~\cite{dalek2013method} use a confirmation methodology to identify URL filtering using, e.g., McAfee SmartFilter and Netsweeper, and detect the use of these technologies in Saudi Arabia, UAE, Qatar, and Yemen.

Nabi~\cite{nabi2013anatomy} uses a publicly available list of blocked websites in Pakistan,  checking their accessibility from multiple networks within the country. Results indicate that censorship varies across websites: some are blocked at the DNS level, while others at the HTTP level.
Furthermore, Verkamp and Gupta~\cite{verkamp2012inferring} detect censorship technologies in 11 countries, mostly using Planet Labs nodes, and discover DNS-based and router-based filtering.
Crandall et al.~\cite{crandall2007conceptdoppler} propose an architecture for maintaining a censorship ``weather report'' about what keywords are filtered over time, while Leberknight et al.~\cite{leberknight2012taxonomy} provide an overview of research on censorship resistant systems and a taxonomy of anti-censorship technologies.

Also, Knockel et al.~\cite{knockel2011three} obtain a built-in list of censored keywords in China's TOM-Skype and run experiments to understand how filtering is operated, while
King et al.~\cite{king2012censorship} devise a system to locate, download, and analyze the content of millions of Chinese social media posts, before the Chinese government censors them. %

Finally, Park and Crandall~\cite{park2010empirical} present results from measurements of the filtering of HTTP HTML responses in China, which is based on string matching and TCP reset injection by backbone-level routers. Xu et al.~\cite{xu2011internet} explore the AS-level topology of China's network infrastructure, and probe the firewall to find the locations of filtering devices, finding that even though most filtering occurs in border ASes, choke points also exist in many provincial networks.

\section{Datasets Description}\label{sec:background}

This section overviews the dataset studied in this paper, and background information on the proxies used for censorship.

\subsection{Data Sources}\label{subsec:sources}
On October 4, 2011, a ``hacktivist'' group called Telecomix announced the release of log files extracted from 7 Syrian Blue Coat SG-9000 
proxies (aka ProxySG)~\cite{leak}. The initial leak concerned 15 proxies but only data from 7 of them was publicly released. As reported by the Wall Street 
Journal~\cite{valentino} and CBS news~\cite{cbs}, Blue Coat openly acknowledged that at least 13 of its proxies were used in Syria, but denied it authorized their sale to the Syrian government~\cite{bluecoat-update}.
These devices have allegedly been used by the Syrian Telecommunications Establishment (STE backbone) to filter and monitor all  connections at a country scale. The data is split by proxy (SG-42, SG-43,$\ldots$, SG-48) and covers two periods: (i) July 22, 23, 31, 2011 (only SG-42), and (ii) August 1--6, 2011 (all proxies). 
The leaked log files are in csv format (comma separated-values) and include 26 fields, such as date, time, filter action, host and URI (more details are given in Section~\ref{subsec:datasets}).

 Given the nature of the dataset, one could question the authenticity of the logs. However,
Blue Coat {\em confirmed} the use of its devices in Syria~\cite{valentino,cbs} and a few findings emerging from the analysis of the logs actually correspond to events and facts that were independently reported before. Also, this leak is not the first censorship-related project carried out by Telecomix. Thus, we are confident that the datasets studied in this paper provide an accurate snapshot of Syrian censorship activities in Summer 2011.

\subsection{Blue Coat SG-9000 Proxies}\label{subsec:appliance}
The Blue Coat SG-9000 proxies perform filtering, monitoring, and caching of Internet traffic, and are typically placed between a monitored network and the Internet backbone. They can be set as {\em explicit} or {\em transparent} proxies: 
the former setting requires the configuration of the clients' browsers, whereas transparent proxies seamlessly intercept traffic (i.e., without clients noticing it), which is the case in this dataset.

Monitoring and filtering of traffic is conducted at the application level. 
Each user request is intercepted and classified as one of the following three labels (as per the \emph{sc-filter-result} field in the logs): 

\begin{compactitem}
 \item \texttt{OBSERVED} -- request is served to the client.
 \item \texttt{PROXIED} -- request has been found in the cache and the outcome depends on the cached value.
 \item \texttt{DENIED} -- request is not served to the client because an exception has been raised (might be redirected).
\end{compactitem}

The classification reflects the action that the proxy needs to perform, rather than the outcome of a filtering process. \texttt{OBSERVED} means that content needs to be fetched from the Origin Content Server (OCS), \texttt{DENIED} that there is no need to contact the OCS, and \texttt{PROXIED} -- that the outcome is in the proxy's cache. 
According to Blue Coat's documentation~\cite{slideshare}, filtering is based on multiple criteria: website categories, keywords, content type, browser type and date/time of day. %
The proxies can also cache content, e.g., to save bandwidth, in the ``bandwidth gain profile'' (see page 193 in~\cite{wikileaks}).

\subsection{Datasets and Notation}\label{subsec:datasets}
Throughout the rest of this paper, our analysis will use the following four datasets:

\begin{enumerate}
\item \textbf{\em Full Logs} (\df): The whole dataset (i.e., extracted from all logs) is composed of 751,295,830 requests.
\item \textbf{\em Sample Dataset} (\ds): Most of the results shown in this paper rely on the full extraction of the relevant data from  \df\ , however, given the massive size of the log files ($\sim$600GB), we \red{sometimes consider a \textit{random} sample covering 4\% of the entire dataset. This dataset (\ds) is only used to illustrate a few results, specifically, for a few summary statistics.} According to standard theory about confidence intervals for proportions (see \cite{StatisticalSignificance91}, Equation 1, Chapter 13.9.2), for a sample size of n = 32M, the actual proportion in the full data set lies in an interval of $\pm$0.0001 around the proportion p observed in the sample with 95\% probability ($\alpha$ = 0.05). 
\item \textbf{\em User Dataset} (\du): Before the data release, Telecomix suppressed user identifiers (IP addresses) by replacing them with zeros. However, for a small fraction of the data (July 22-23), user identifiers were replaced with the hash of the IP addresses, thus making user-based analysis possible. %
\item \textbf{\em Denied Dataset} (\dd): This dataset contains all the requests that resulted in exceptions (\emph{x-exception-id} $\neq$ `-'), and hence are not served to the user.
\end{enumerate}

In Table~\ref{tab:dataStat1}, we report, for each dataset, the number of requests in it, corresponding dates, and number of proxies.
Then, in Table~\ref{tab:field_description}, we list a few fields from the logs that constitute the main focus of our analysis.
The \emph{s-ip} field logs the IP address of the proxy that processed each request, which is in the range 82.137.200.42 -- 48. Throughout the paper we refer to the proxies as SG-42 to SG-48, according to the suffix of their IP address.
The \emph{sc-filter-result} field indicates whether the request has been served to the client. In the rest of the paper, we consider as \emph{denied} all requests that have not been successfully served to the client by the proxy, including requests generating network errors as well as requests censored based on policy. To further classify a denied request, we rely on the \emph{x-exception-id} field: all denied requests which either raise  \policydenied or \policyredirect flags are considered as \emph{censored}. %

\begin{table}[t]
\centering
\small
\begin{tabular}{|l|c|c|c|}
	\hline
	 {\bf Dataset} & {\bf \# Requests} & {\bf Period}  & {\bf \# Proxies} \\ 
	\hline 	
	\multirow{2}{*}{\bf Full} & \multirow{2}{*}{751,295,830} &  July 22-23,31, 2011 & \multirow{2}{*}{7} \\
	& & August 1-6, 2011 &  \\ 
\hline
	\multirow{2}{*}{\bf Sample (4\%)} & \multirow{2}{*}{32,310,958} &  July 22-23, 2011 & \multirow{2}{*}{7} \\
	& & August 1-6, 2011 &  \\ 
\hline
	{\bf User} & 6,374,333	 & July 22-23 2011 & 1  \\ 
\hline
	\multirow{2}{*}{\bf Denied} & \multirow{2}{*}{47,452,194} & July 22-23,31, 2011 & \multirow{2}{*}{7} \\
	& & August 1-6, 2011  &  \\ 
\hline
\end{tabular} 
\caption{Datasets description.}\label{tab:dataStat1}
\end{table}

\begin{table}[t]
\small
\centering
\begin{tabular}{|l|p{6cm}|} 	
	\hline
 {\bf Field name} &  {\bf Description} \\
 \hline
   {\em cs-host} &  Hostname or IP address (e.g., \url{facebook.com})\\  \hline
   {\em cs-uri-scheme} & Scheme used by the requested URL (mostly HTTP) \\ \hline
   {\em cs-uri-port} & Port of the requested URL  \\ \hline
   {\em cs-uri-path} & Path of the requested URL (e.g., /home.php) \\ \hline
   \multirow{2}{*}{\em cs-uri-query} & Query of the requested URL  \\
    & (e.g., ?refid=7\&ref=nf\_fr\&\_rdr) \\  \hline
   {\em cs-uri-ext} & Extension of the requested URL (e.g., php, flv, gif, ...) \\  \hline
   {\em cs-user-agent} & User agent (from request header)\\ \hline
      \multirow{2}{*}{\em cs-categories} & Categories to which the requested URL has been classified (see Section~\ref{sec:general_characteristics} for details)\\ \hline
    {\em c-ip} & Client's IP address (removed or anonymized)\\ \hline
   \multirow{2}{*}{\em s-ip} & The IP address of the proxy that processed the client's request \\ \hline
   \multirow{2}{*}{\em sc-status} & Protocol status code from the proxy to the client (e.g., `200' for OK) \\ \hline
   {\em sc-filter-result} & Content filtering result: \texttt{DENIED}, \texttt{PROXIED}, or \texttt{OBSERVED}  \\ \hline
   \multirow{2}{*}{\em x-exception-id} & Exception raised by the request (e.g., \policydenied, \emph{dns\_error}). Set to `-' if no exception was raised. \\
   \hline
\end{tabular}
\caption{Description of a few relevant fields from the logs.}\label{tab:field_description}
\end{table} 

Finally, we observe some inconsistencies in the requests that have a \emph{sc-filter-result} value set to \texttt{PROXIED} with no exception. 
When looking at requests similar to those that are \texttt{PROXIED} (e.g., other requests from the same user accessing the same URL), some are consistently denied, while others are sometimes or always allowed.
Since \texttt{PROXIED} requests only represent a small portion of the analyzed traffic ($<0.5\%$), we treat them like the rest of the traffic and classify them according to the \emph{x-exception-id}. However, where relevant, we refer to them explicitly to distinguish them from the \texttt{OBSERVED} traffic.

In summary, throughout the rest of the paper, we use the following request classification:
\begin{compactitem} 
\item \textit{Allowed} (\emph{x-exception-id} = `-'): a request that is allowed and served to the client (no exception raised).
\item \textit{Denied} (\emph{x-exception-id} $\neq$ `-'): a request that is not served to the client, either because of a network error or due to censorship. Specifically:
\begin{compactitem}
	\item \textit{Censored} (\emph{x-exception-id} $\in$ \{\policydenied, \emph{policy\_redirect}\}): a \emph{denied} request that is censored based on censorship policy.
	\item \textit{Error} (\emph{x-exception-id} $\not\in$ \{`-', \policydenied, \emph{policy\_redirect}\}): a \emph{denied} request not served to the client due to a network error.
\end{compactitem}	
\item \textit{Proxied} (\emph{sc-filter-result} = \texttt{PROXIED}): a request that  does not need further processing, as  the response is in the cache (i.e., the result depends on a prior computation). The request can be either allowed or denied, even if \emph{x-exception-id} does not indicate an exception. 
\end{compactitem}

\subsection{Ethical Considerations}\label{subsec:ethics}

Even though the dataset studied in this paper is publicly available, we are obviously aware of its sensitivity. Thus, we enforced a few mechanisms to safeguard  privacy of Syrian Internet users. We encrypted all data (and backups) at rest and did not re-distribute the logs. We never obtained or extracted users' personal information, and we only analyzed aggregated traffic statistics.
While it is out of the scope of this paper to further discuss the ethics of using ``leaked data'' for research purposes (see \cite{bonneau} for a detailed discussion), we argue that analyzing logs of filtered traffic, as opposed to probing-based measurements, provides an accurate view for a large-scale and comprehensive analysis of censorship.\footnote{Note that we obtained the approval of INRIA's and NICTA's Institutional Review Board (IRB) to publish results of this work.}

We acknowledge that our work may be beneficial to entities on 
either side of censorship. However, our analysis helps understand the technical aspects of an actual censorship ecosystem. Our methodology exposes its
underlying technologies, policies, as well as its strengths and weaknesses, and can thus facilitate the design of censorship-evading tools.

\section{A Statistical Overview of\\ Censorship in Syria}
\label{sec:general_characteristics}

Aiming to provide an overview of Internet censorship in Syria,
our first step is to compare the statistical distributions of the different classes of traffic (as defined in Section~\ref{subsec:datasets}), and also look at domains, TCP/UDP ports, website categories, and HTTPS traffic.
Unless explicitly stated otherwise, the results presented in this section are based on the full dataset denoted as \df\ (see Section~\ref{subsec:datasets}).

\descr{Traffic distribution.} We start by observing the ratio of the different classes of traffic. For each of the datasets \df, \ds, \du\ and \dd, 
Table~\ref{tab:exception} reports how many requests are allowed, proxied, denied, or censored. 
In \df, more than 93\% of the requests are allowed, and less than 1\% of them are censored due to policy-based decisions. The number of censored requests seems relatively low compared to the  number of allowed requests. Note, however, that these numbers are skewed because of the request-based logging mechanism, which ``inflates'' the volume of allowed traffic; a single access to a website may trigger a large number of requests (e.g., for the HTML content, accompanying images, scripts, tracking websites and so on) that will be logged, whereas a \emph{denied} request (either because it has been censored or due to a network error) only generates one log entry. 
Finally, note that only a small fraction of requests are proxied (0.47\% in \df). The breakdown of \emph{x-exception-id} values within the proxied requests resembles that of the overall traffic. %

\begin{table*}[t!]
\centering
\small
		\begin{tabular}{|l|l|l|c|r c|r c|r c|r c|}
			\hline
			\multicolumn{2}{|l|}{\bf sc-filter-result} & \bf{x-exception-id} & \bf{Class} & \multicolumn{2}{c|}{\bf Full ($D_{full}$)}  & \multicolumn{2}{c|}{\bf Sample ($D_{sample}$)} &
			\multicolumn{2}{c|}{\bf User ($D_{user}$)} & \multicolumn{2}{c|}{\bf Denied ($D_{denied})$}\\
			\hline
			\multicolumn{2}{|l|}{\texttt{OBSERVED}} & -- & \emph{Allowed}  &  700,606,503  & (93.25\%) & 30,140,158 & (93.28\%) & 6,038,461 & (94.73\%) & -- & -- \\
			\hline
			\multicolumn{2}{|l|}{\texttt{PROXIED}} & (total) & \emph{Proxied} &  3,504,485  & (0.47\%) & 151,554 & (0.47\%) & 26,541 & (0.42\%) & 267,354& (0.56\%) \\ %
			\hline
			\multicolumn{2}{|l|}{\texttt{DENIED}} & (total) & \emph{Denied} &  47,184,840  & (6.28\%) &  2,019,246 & (6.25\%) & 309,331 & (4.85\%) & 47,184,840 & (99.44\%) %
			\\ \cline{2-12} 
			& \multicolumn{2}{l|}{tcp\_error} &\multirow{9}{*}{\emph{Error}} &  21,499,871  & (2.86\%) &  947,083 & (2.93\%) & 54,073 & (0.85\%) & 21,499,871 & (45.30\%)  \\
			& \multicolumn{2}{l|}{internal\_error} & &  14,720,952  & (1.96\%) & 636,335 & (1.97\%) & 198,058 & (3.11\%) & 14,720,952 & (31.02\%)  \\
			& \multicolumn{2}{l|}{invalid\_request} & &  2,668,217  & (0.36\%) & 115,297 & (0.36\%) & 36,292 & (0.57\%) & 2,668,217 & (5.62\%)  \\
			& \multicolumn{2}{l|}{unsupported\_protocol} & &  719,189  & (0.10\%) & 28,769 & (0.09\%) & 1,348 & (0.02\%) & 719,189 & (1.51\%)  \\ 
			& \multicolumn{2}{l|}{dns\_unresolved\_hostname} & &   141,558  & (0.02\%) & 6,247 & (0.02\%) & 3,856 & (0.06\%) & 141,558 & (0.30\%)  \\

			& \multicolumn{2}{l|}{dns\_server\_failure} &  &  58,401  & (0.01\%) & 2,235 & (0.01\%) & 396 & (0.01\%) & 58,401 & (0.12\%)  \\ 
			& \multicolumn{2}{l|}{unsupported\_encoding} & &  269  & (0.00\%) & 6 & (0.00\%) & 0 & (0.00\%) & 269 & (0.00\%)  \\ 
			& \multicolumn{2}{l|}{invalid\_response} & & 8 & (0.00\%) & 1 & (0.00\%) & 2 & (0.00\%) & 8 & (0.00\%)  \\  \cline{2-12} 

			& \multicolumn{2}{l|}{policy\_denied} & \multirow{2}{*}{\emph{Censored}} &   7,374,500  & (0.98\%) & 283,197 & (0.88\%) & 15,306 & (0.24\%) & 7,374,500 & (15.54\%)  \\ 
			& \multicolumn{2}{l|}{policy\_redirect} & &  1,875  &(0.00\%) & 76 & (0.00\%) & 0 & (0.00\%) & 1,875 & (0.04\%)  \\ 

		\hline
		\end{tabular}
\caption{Statistics of different decisions and exceptions in the three datasets in use.}
\label{tab:exception}
\end{table*}

\descr{Denied traffic.}
Proxies also log requests that have been denied due to network errors: this happens for less than 6\% of the requests in our sample.
The inability of the proxy to handle the request (identified by the {\em x-exception-id} field being set to \emph{internal\_error}) accounts for 31.15\% of the overall denied traffic. Although this could be considered  censorship (no data is received by the user), these requests do not actually trigger any policy exception and are not the result of policy-based censorship. TCP errors, typically occurring during the connection establishment between the proxy and the target destination, represent more than 45\% of the denied traffic.  Other errors include DNS resolving issues (0.41\%), invalid HTTP request or response formatting (5.65\%), and unsupported protocols (1.46\%). The remaining 15.33\% of denied traffic represent the actual censored requests, which the proxy flags as denied due to policy enforcement.

\begin{figure}[t!]
 \begin{center}
 \includegraphics[width=0.43\textwidth]{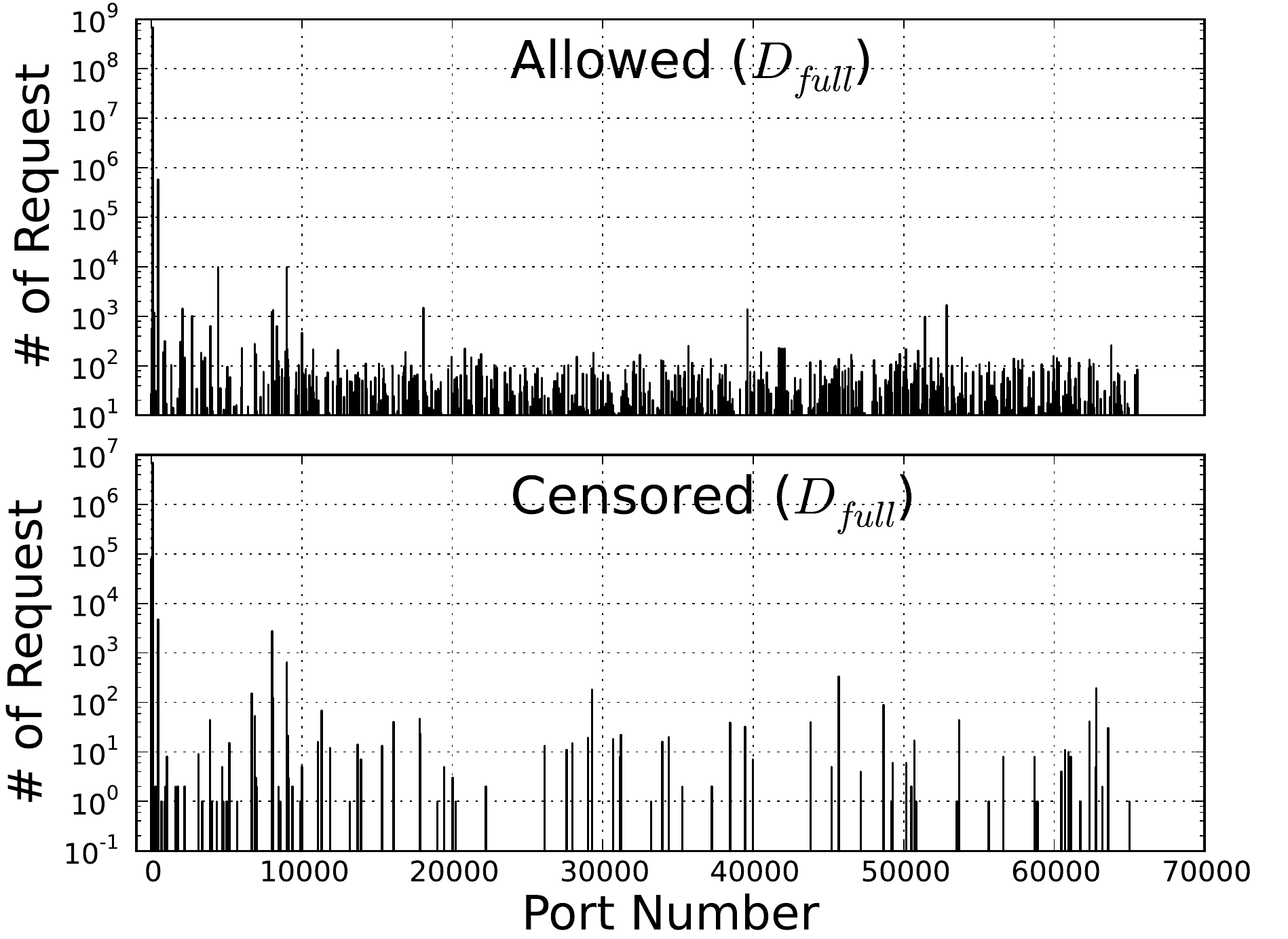}
\vspace{-0.15cm}
\caption{Destination port distributions of allowed and censored traffic ($D_{full}$).}
\label{fig:port_distribution_sample}
\end{center}
\end{figure}

\descr{Ports.} We also look at the traffic distribution by port number for both allowed and censored traffic (in \df). We report it in Fig.~\ref{fig:port_distribution_sample}. Ports 80 and 443 (HTTPS) represent the majority of censored content. Port 9001 (usually associated with Tor servers) is ranked third in terms of blocked connections. We discuss Tor traffic in more detail in Section~\ref{sec:tor}.

\begin{figure}[b!]
\centering
 \includegraphics[height=0.226\textheight]{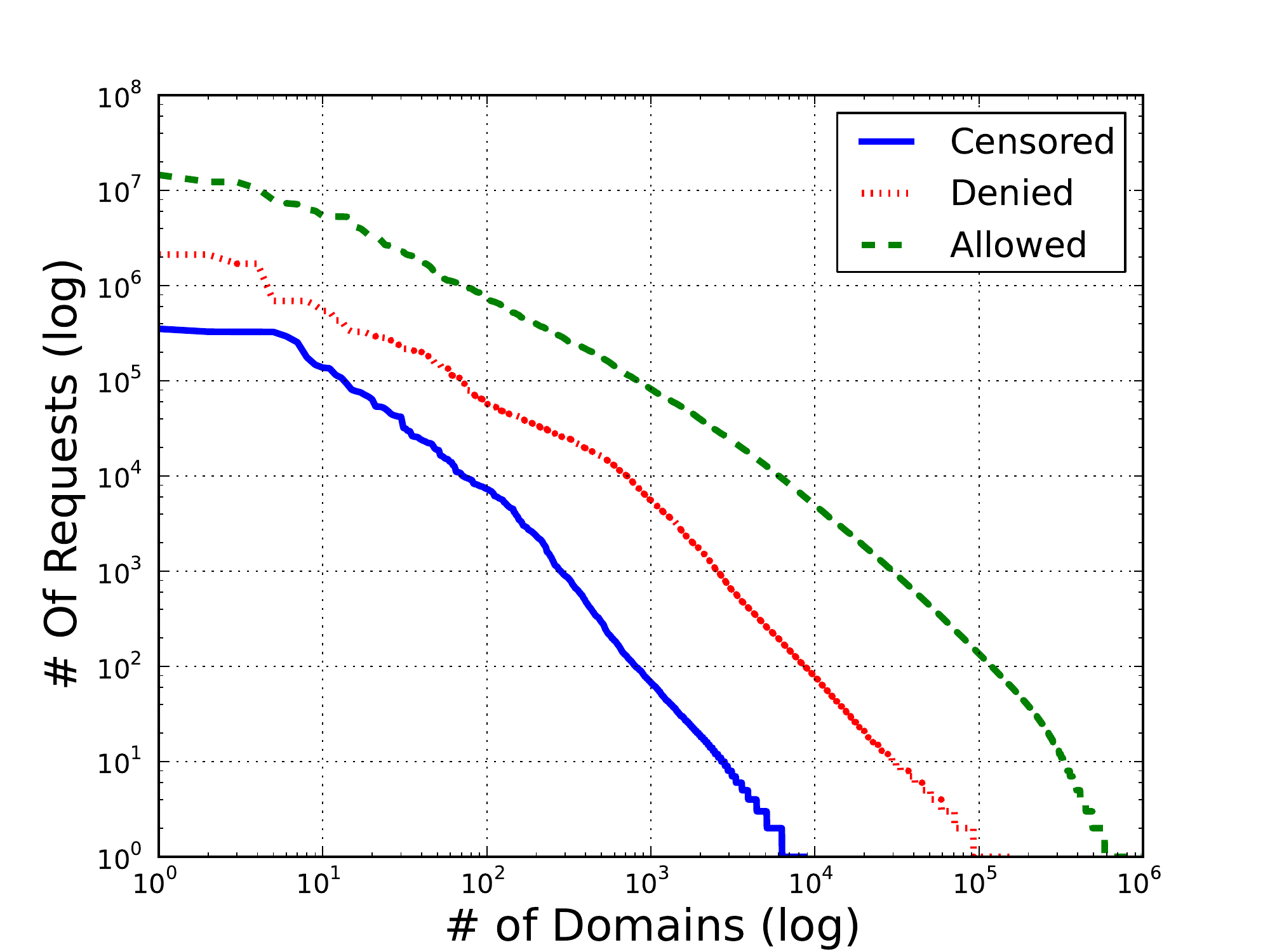}
\caption{Distribution of \# requests per unique domain (\df).}
\label{fig:cdf_host}
\end{figure}

\descr{Domains.} Next, we analyze the distribution of the number of requests per unique domain. 
Fig.~\ref{fig:cdf_host} presents our findings. The y-axis (log-scale) represents the number of (allowed/denied/censored) requests, while each point in the x-axis (also log-scale) represents the number of domains receiving such a number of requests. Unsurprisingly, the curves indicate a power law distribution.  We observe that a very small fraction of hosts ($10^{-5}$ for the allowed requests) are the target of between few thousands to few millions requests, while the vast majority are the destination of only few requests. Allowed traffic is at some point one order of magnitude bigger than denied traffic, this happens for at least two reasons: (i) allowed requests target highly popular websites (e.g., Google and Facebook), and  (ii) an allowed request is potentially followed up by additional requests to the same domain, whereas a denied request is not.

In Table \ref{tab:top10domainsAllowedCensored}, we report the top-10 allowed (resp., censored) domains in \df. Unsurprisingly, \url{google.com} and its associated static/tracking/advertisement components represent nearly 15\% of the total allowed requests. Other well-ranked domains include \url{facebook.com} (and its associated CDN service, \url{fbcdn.net}) and \url{xvideos.com} (a pornography-associated  website). 
The top-10 censored domains exhibit a very different distribution: \url{facebook.com} (and \url{fbcdn.net}), \url{metacafe.com} (a popular user-contributed video sharing service)  and \url{skype.com} account for more 
than 43\% of the overall censored requests. Finally, observe that websites like Facebook and Google are present both in the censored and the allowed traffic, since the policy-based filtering  may  depend on the actual content the user is fetching rather than the website, as we will explain in Section~\ref{sec:social_media}.
%

%
%
%
%
%
%
%
%
%
%
%
%
%
%
%
%
%
%
%
%
%
%
%
%
%
%
%
%
%
%

\begin{table}[t!]
	\centering
	\vspace{0.1cm}
    \resizebox{0.99\columnwidth}{!}{
		\begin{tabular}{|c|c|c|c|}
		\hline
			\multicolumn{2}{|c|}{\bf Allowed domains} & \multicolumn{2}{|c|}{\bf Censored domains}  \\ \hline
			{\bf Domain} & {\bf \# Requests (\%)} & {\bf Domain} & {\bf \# Requests (\%)} \\ \hline
			google.com & 50.36M (7.19\%) & facebook.com & 1.62M (21.91\%) \\
			xvideos.com & 23.42M (3.34\%) & metacafe.com & 1.28M (17.33\%)\\
			gstatic.com & 23.10M (3.30\%) & skype.com & 503,932 (6.83\%)\\
			facebook.com & 17.83M (2.54\%) & live.com & 441,408 (5.98\%)\\
			microsoft.com & 16.64M (2.38\%) & google.com & 420,862 (5.71\%) \\
			fbcdn.net & 16.46M (2.35\%) & zynga.com & 379,170 (5.14\%)\\
			windowsupdate.com & 15.43M (2.20\%) & yahoo.com & 369,948 (5.02\%) \\
			google-analytics.com & 12.38M (1.77\%) & wikimedia.org & 306,994 (4.16\%)\\
			doubleclick.net & 11.19M (1.60\%) & fbcdn.net & 264,512 (3.59\%)\\
			msn.com & 11.01M (1.57\%) & ceipmsn.com & 135,134 (1.83\%)\\
		\hline
		\end{tabular}
		}
\caption{Top-10 Domains (allowed and censored) in  \df.}
	\label{tab:top10domainsAllowedCensored}
\end{table}

\descr{Categories.} The Blue Coat proxies support filtering according to URL categories. This categorization can be done using a local database, or using Blue Coat's online filtering tool.\footnote{\url{http://sitereview.bluecoat.com/categories.jsp}} However, according to Blue Coat's representatives \cite{valentino}, the online services are not accessible to the Syrian proxy servers, and apparently the Syrian proxy servers are not using a local copy of this categorization database.
Indeed, the \emph{cs-categories} field in the logs, which records the URL categories, contains only one of two values: one value associated with a default category (named ``unavailable'' in five of the proxies, and ``none'' in the other two), and another value associated with a custom category targeted at Facebook pages (named ``Blocked sites; unavailable'' in five of the proxies, and ``Blocked sites'' in the other two), which is discussed in more details in Section~\ref{sec:social_media}. 

\begin{figure}[t!]
 \begin{center}
 \includegraphics[width=0.4\textwidth]{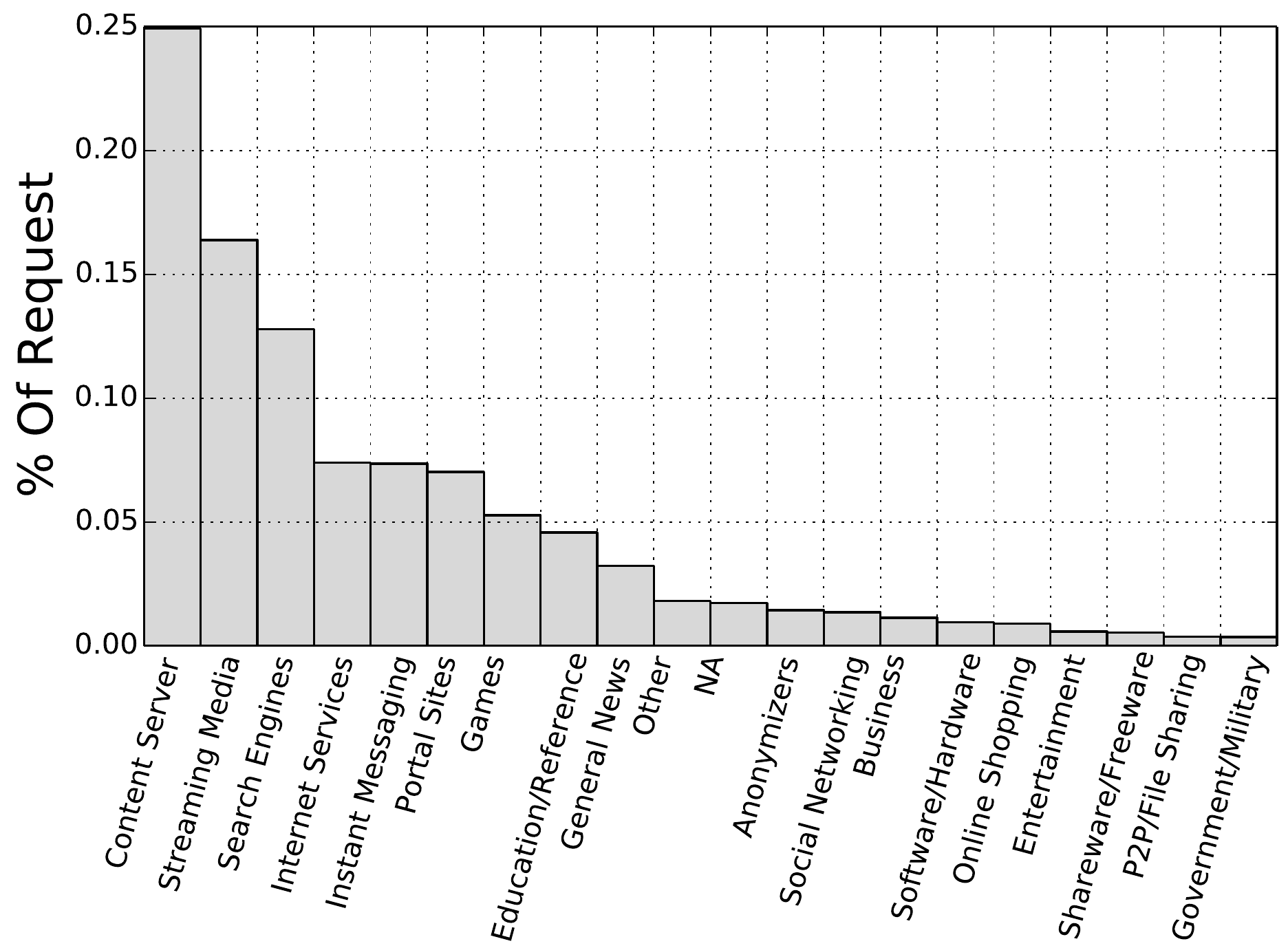}
\caption{Category distribution of censored traffic ($D_{sample}$). Categories obtained from McAfee's TrustedSource, `Other' is used for categories with less than 1K requests.}
\label{fig:ctg_dictribution_censored_sample}
\end{center}
\end{figure}

Due to the absence of URL categories, we use McAfee's TrustedSource tool (\url{www.trustedsource.org}) to characterize the censored websites. Fig.~\ref{fig:ctg_dictribution_censored_sample} shows the distribution of the censored requests across the different categories. The ``Content Server'' category ranks first, with more than 25\% of the blocked requests (this category mostly includes CDNs that host a variety of websites, such as \url{cloudfront.net}, \url{googleusercontent.com}). 
``Streaming Media'' are next, hinting at the intention of the censors to block video sharing. ``Instant Messaging'' (IM) websites, as well as ``Portals Sites'', are also highly blocked, possibly due to their role in coordination of social activities and protests. Note that both Skype and live.com IM services are always censored and belong to the top-10 censored domains. However, surprisingly, both ``News Portals'' and ``Social Networks'' rank relatively low: as we explain in Section~\ref{sec:social_media}, censorship only blocks a few well-targeted social media pages. Finally, categories like ``Games'' and ``Education/Reference'' are also occasionally blocked.

\descr{HTTPS traffic.} 
The number of HTTPS requests is a few orders of magnitude lower than that of HTTP requests. HTTPS accounts for 0.08\% of the overall traffic and only a small fraction (0.82\%) is censored (\ds~dataset). 
It is interesting to observe that, in 82\% of the censored traffic, the destination field indicates an IP address rather than a domain, and such an IP-based blocking occurs at least for two main reasons: (1) the IP address belongs to an Israeli AS, or (2) the IP address is associated with an Anonymizer service. 
The remaining part of the censored HTTPS traffic actually contains a hostname: 
this is possible due to the use of the HTTP CONNECT method, which allows the proxy to identify both the destination host and the user agent (for instance, all connections to Skype servers are using the CONNECT method, and thus the proxy can censor requests based on the \url{skype.com} domain). %

According to the Electronic Frontier Foundation, the Syrian Telecom Ministry has launched man in the middle (MITM) attacks against the HTTPS version of Facebook.\footnote{\url{https://www.eff.org/deeplinks/2011/05/syrian-man-middle-against-facebook}} While Blue Coat proxies indeed support interception of HTTPS traffic,\footnote{\url{https://kb.bluecoat.com/index?page=content&id=KB5500}} we do not identify any clear sign of such an activity. For instance, the values of fields such as \emph{cs-uri-path}, \emph{cs-uri-query} and \emph{cs-uri-ext}, which would have been available to the proxies in a MITM attack, are not present in HTTPS requests. However, also note that, by default, the Blue Coat proxies use a separate log facility to record SSL traffic,\footnote{See \url{https://bto.bluecoat.com/doc/8672}, page 22.} so it is possible that this traffic has been recorded in logs that were not obtained by Telecomix.

\descr{User-based analysis.} Based on the \du dataset, which comprises  the logs of proxy SG-42 from July 22-23, we analyze user behavior with respect to censorship. We assume that each unique combination of {\em c-ip} (client IP address) and {\em cs-user-agent} designates a unique user. 
This means that a user connecting from several devices with different IP addresses (or a single device with different browsers) is not considered as a single users. Conversely, users behind NAT using browsers with identical user-agent are counted as one user. However, this combination of fields provides the best approximation of unique users within the limits of the available data~\cite{yen:tracking}.

\begin{figure}[t!]
  \begin{center}
  \subfigure[]{
           \centering
			\includegraphics[width=0.33\textwidth]{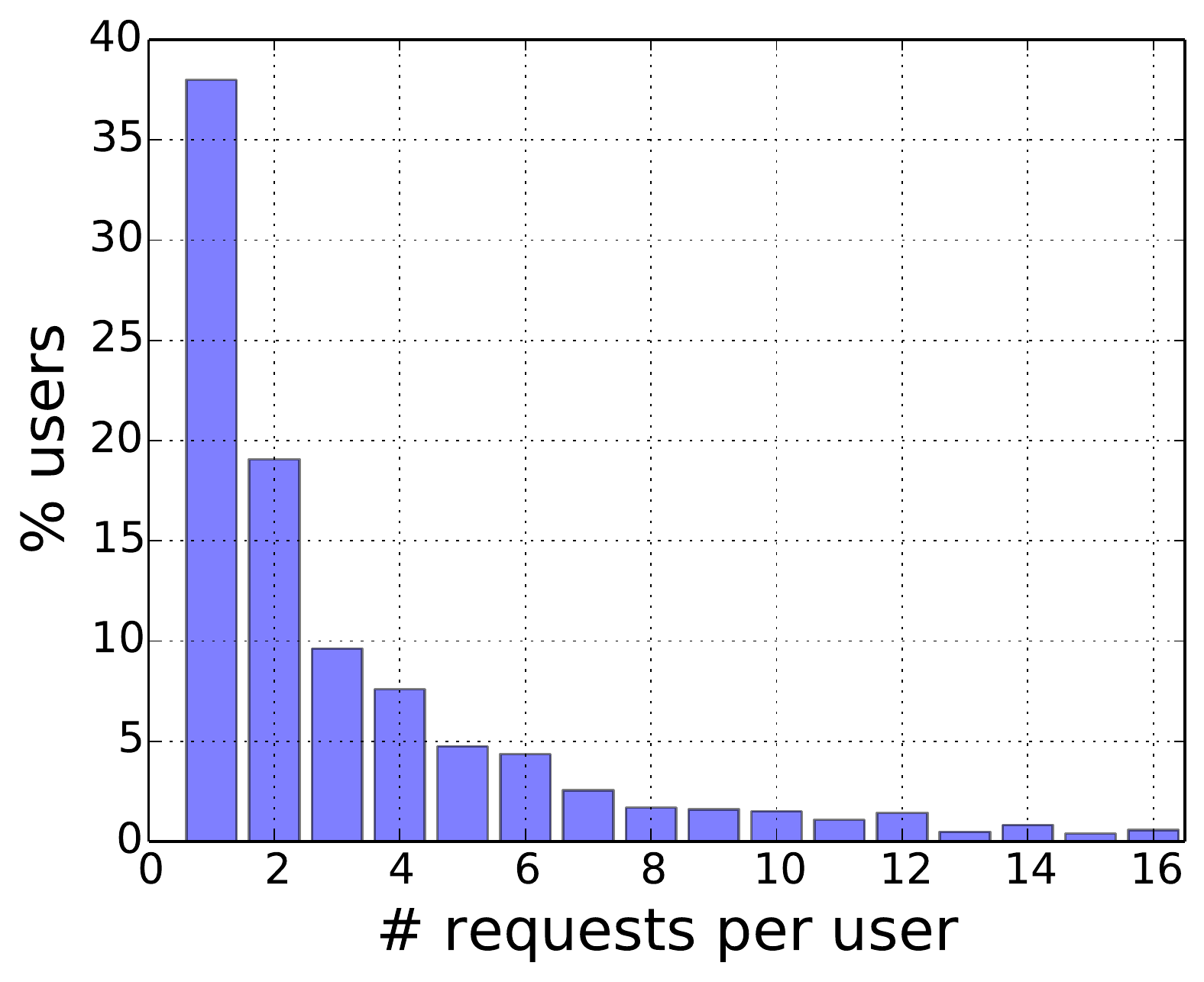}
			\label{fig:userCensoredNumber}
        }
   \subfigure[]{
            \centering
			\includegraphics[width=0.33\textwidth]{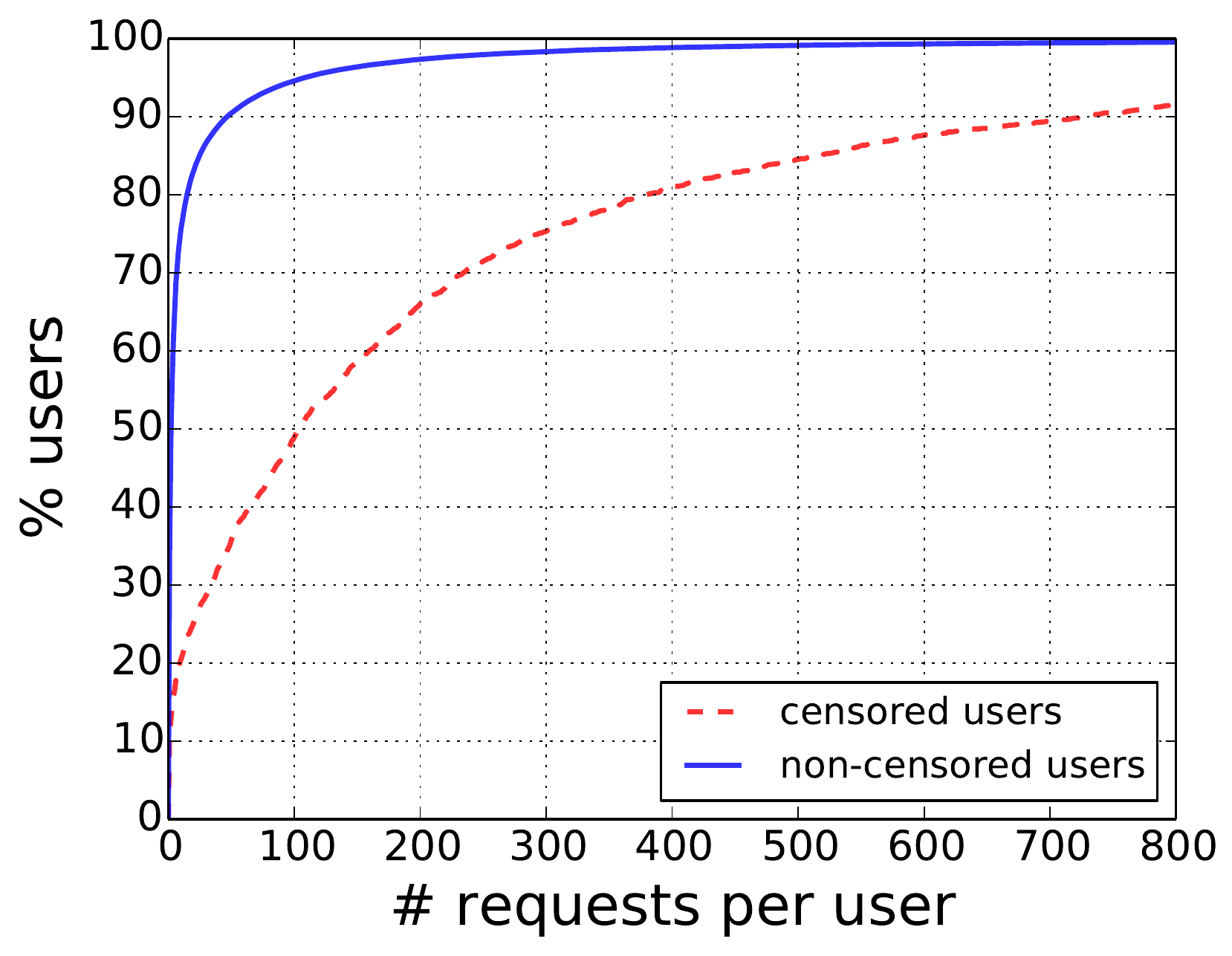}
			
			\label{fig:userAllowedNumber}
        }
        \vspace{-0.4cm}
\caption{(a) Number of censored requests per user in \du; (b) The distribution of the overall number of requests per user (both allowed and denied), for censored and non-censored users.}
        \vspace{-0.7cm}

	\end{center}
\end{figure}

We identify 147,802 total users in \du. 2,319 (1.57\%) of them generate at least one request that is denied due to censorship. 
%
%
%
%

Fig.~\ref{fig:userAllowedNumber} shows the distribution of the number of overall requests per user, for both non-censored and censored users, where a censored user is defined as a user for whom at least one request was censored. We found that the censored users are more active than non-censored users, observing approximately 50\% of the censored users have sent more than 100 requests, while only 5\% of non-censored users show the same level of activity. As we discuss in Section~\ref{subsec:cat_string_ip_censorship}, many requests are censored since they happen to contain a blacklisted keyword (e.g., {\em proxy}), even though they may not be actually accessing content that is the target of censorship. Since active users are more likely to encounter URLs that contain such keywords, this may explain the correlation between the user level of activity and being censored. We also observe that in some cases the user agent field refers to a software repeatedly trying to access a censored page (e.g., \url{skype.com}), which augments the user's activity.

\descr{Summary.}
Our measurements have shown that only a small fraction (<1\%) of the traffic is actually censored. Most requests are either allowed (93.28\%) or denied due to network errors (5.37\%). Censorship targets mostly HTTP content, but several other services are also blocked. Unsurprisingly, most of the censorship activity targets websites that support user interaction (e.g., Instant Messaging and social networks). 
A closer look at the top allowed and censored domains shows that some hosts are in both categories, thus hinting at a more sophisticated censoring mechanism, which we explore in the next sections.
Finally, our user-based analysis has shown that only a small fraction of users are directly affected by censorship.

\section{Understanding the Censorship}
\label{sec:understanding_censorhip}

This section aims to understand the way the Internet is filtered in Syria.
We analyze censorship's temporal characteristics and compare the behavior of different proxies. Then, we study {\em how} the requests are filtered and infer characteristics on which censorship policies are based. 

\subsection{Temporal Analysis}\label{subsec:temporal}

\begin{figure}[t!]
  \centering
  \subfigure[]{ 
            \includegraphics[width=0.435\textwidth]{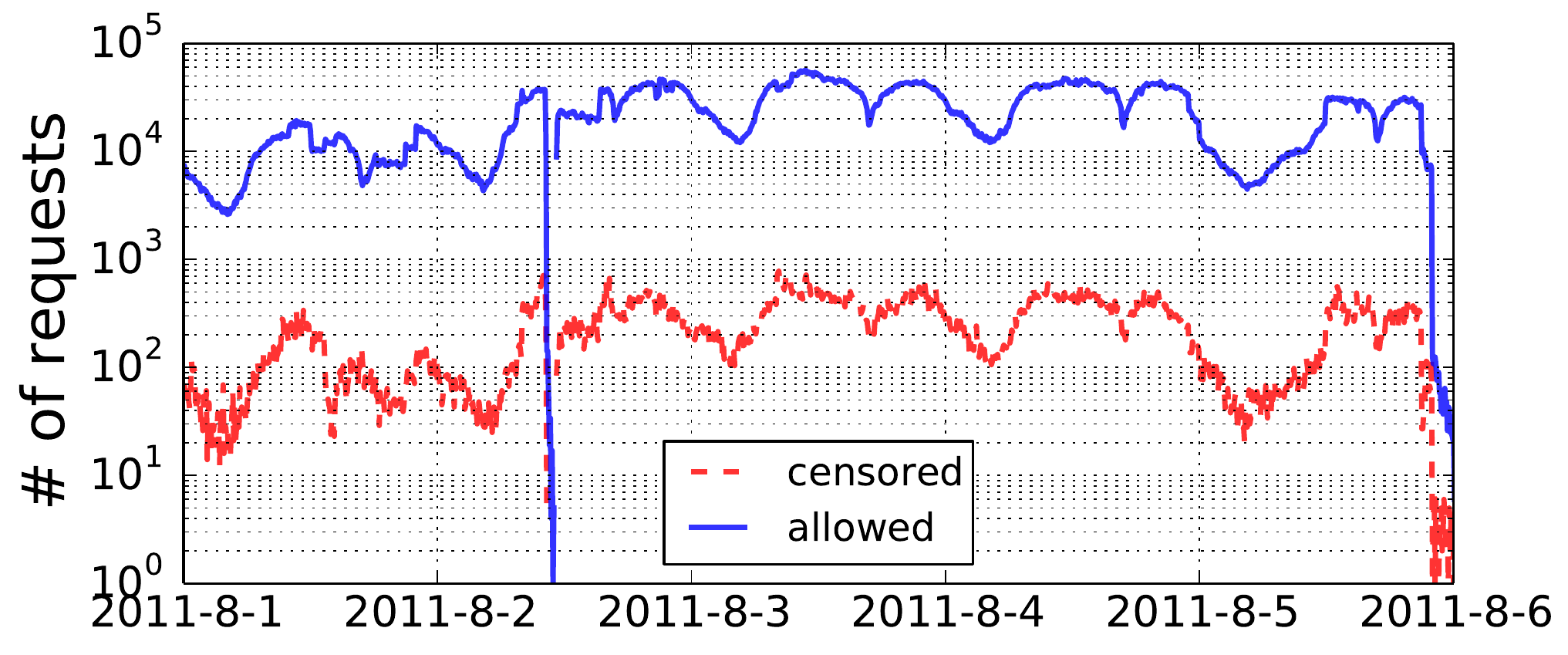}
 			\label{fig:tempAbsolute}
        }
   \subfigure[]{%
            \includegraphics[width=0.435\textwidth]{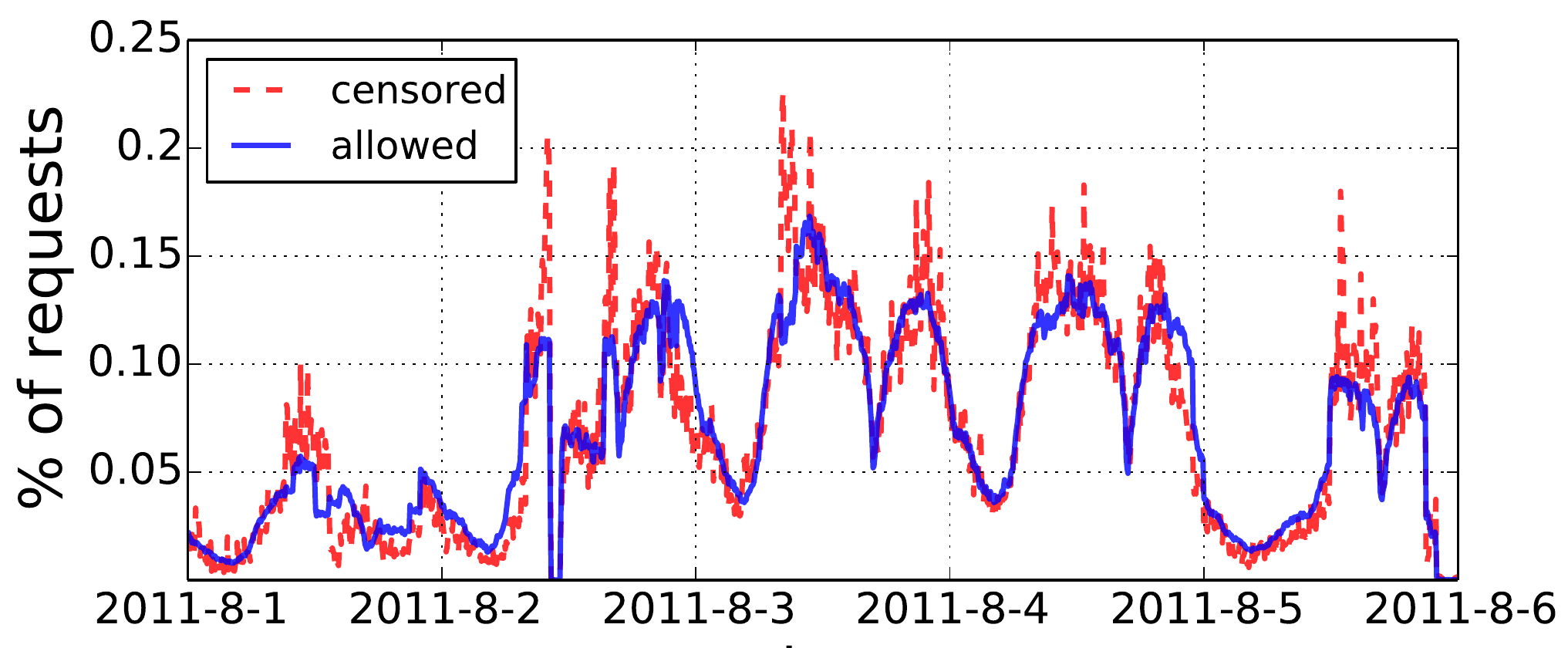}
 			\label{fig:tempNormalized}
        }
        \vspace{-0.2cm}
\caption{Censored and allowed traffic over 5 days (absolute/normalized). }
	\label{fig:temp5days}
\end{figure}

We start by looking at how the traffic volume of both censored and allowed traffic changes over time (5 days), with 5-minute granularity. The corresponding time-series are reported in Fig.~\ref{fig:tempAbsolute}:
as expected, they roughly follow the same patterns, with an increasing volume of traffic early mornings, followed by a smooth lull 
during afternoons and nights.
To evaluate the overall variation of the censorship activity, we show in Fig.~\ref{fig:tempNormalized} the temporal evolution of the number of censored (resp., allowed) requests at specific times of the day, {\em normalized} by the total number of censored (resp., allowed) requests. Note that the two curves are not comparable, but illustrate the relative activity when considering the overall nature of the traffic over the observation period. The relative censorship activity exhibits a few peaks, with a higher volume of censored content on particular periods of time. There are also two sudden ``drops'' in both allowed and censored requests, which might be correlated to some protests that day.\footnote{See
\url{http://www.enduringamerica.com/home/2011/8/3/syria-and-beyond-liveblog-the-sights-and-sounds-of-protest.html}.}  
There is a visible reduction in traffic from  Thursday afternoon (August 4) to Friday (August 5), consistent with press reports of Internet connections being slowed almost every Friday ``when the big weekly protests are staged''~\cite{RWB}.

\begin{figure} [t!]
\centering
 \includegraphics[width=0.42\textwidth]{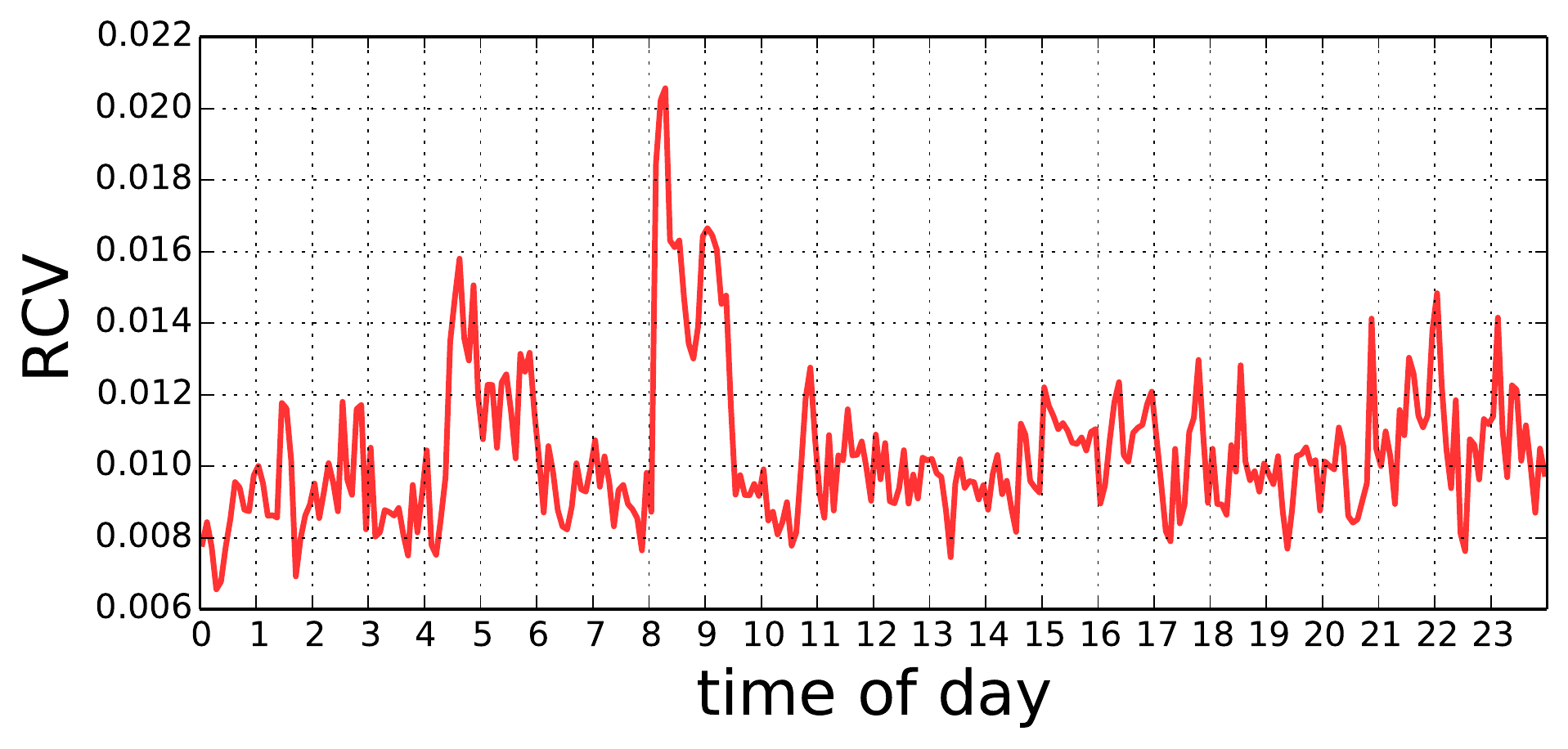}
\vspace{-0.2cm}
\caption{Relative Censored traffic Volume (RCV) for August 3 (in $D_{sample}$) as a function of time.}\label{fig:rcv}
\end{figure}

To further study the activity peaks, we zoom in on one specific day (August 3) that has a particularly high volume of \red{censored traffic}.  Let RCV  ({\em Relative Censored traffic Volume}) be the ratio between
the number of censored requests at a time frame (with a 5-minute granularity) and the total number of requests received on the same time frame; in
Fig.~\ref{fig:rcv}, we plot RCV as a function of the time of day. 
There are a few sharp increases in \red{censored traffic}, with the fraction of censored content increasing from 1\% to 2\% of the total traffic around 8am, while, around 9.30am, the RCV variation exhibits a sudden decay. A few other peaks are also observed early morning (5am) and evening (10pm).
 
We then analyze the distribution of censored content between 8am and 9.30am on August 3 and, in  Table~\ref{tab:top10domains}, we report the top-10 censored domains during this period and the adjacent ones (as well as the corresponding percentage of censored volume each domain represents). 
It is evident that \url{skype} is being heavily blocked (up to 29\% of the censored traffic), probably due to the protests that happened in Syria on August 3, 2011. However, 
9\% of the requests to Skype servers are related to update attempts (for Windows clients) and all of them are denied. There is also an unusually higher number of requests to MSN live messenger service (through \url{msn.com}), thus suggesting
that the censorship activity peaks \red{might perhaps be correlated} to high demand targeting Instant Messaging software websites.\footnote{Very similar results also occur for other periods of censorship activity peaks.} In conclusion, censorship peaks might be due to sudden higher volumes of traffic targeting Skype and MSN live messenger websites, which are being systematically censored by the proxies.

\begin{table}[t!]
	\centering
	\resizebox{0.99\columnwidth}{!}{
    \begin{tabular}{|c| c| c | c | c | c| }
    \hline
    \multicolumn{2}{|c }{\bf 6am - 8am} & \multicolumn{2}{ |c|}{\bf 8am - 10am} & \multicolumn{2}{ c| }{\bf 10am -12pm}\\
    \cline{1-6} 
    {\bf Domain} & {\bf \%} & {\bf Domain} & {\bf \%} &  {\bf Domain} & {\bf \%}\\ \hline
    metacafe.com & 20.4\% &  skype.com & 29.24\%  &  facebook.com & 22.47\% \\ \hline
    trafficholder.com & 16.87\% &  facebook.com & 19.45\%  &  metacafe.com & 18.56\% \\ \hline
    facebook.com & 15.08\%  &  live.com & 9.59\%  &  live.com & 11.93\% \\ \hline
    google.com & 8.15\% &  metacafe.com & 7.59\%  &  skype.com & 11.79\% \\ \hline
    yahoo.com & 6.43\%  &  google.com & 6.76\%  &  google.com & 6.81\% \\ \hline
    zynga.com & 5.14\% &  yahoo.com & 3.57\%  &  zynga.com & 3.43\% \\ \hline
    live.com & 3.04\% &  wikimedia.org & 2.47\%  &  ceipmsn.com & 2.38\% \\ \hline
	conduitapps.com & 1.45\% &  trafficholder.com & 2.06\%  &  mtn.com.sy & 2.13\% \\ \hline
	all4syria.info & 1.44\%  &  dailymotion.com & 1.58\%  &  panet.co.il & 1.02\% \\ \hline
	hotsptshld.com & 1.18\% &  conduitapps.com & 1.11\%  &  bbc.co.uk & 0.91\% \\ \hline
    \end{tabular}
    }
\caption{Top censored domains, August 3, 6am-12pm.}
	     \label{tab:top10domains}
\end{table}

\subsection{Comparing different proxies}\label{subsec:inconsistencies}
\begin{figure}[t!]
  \begin{center}
  \subfigure{ 
            \includegraphics[width=0.45\textwidth]{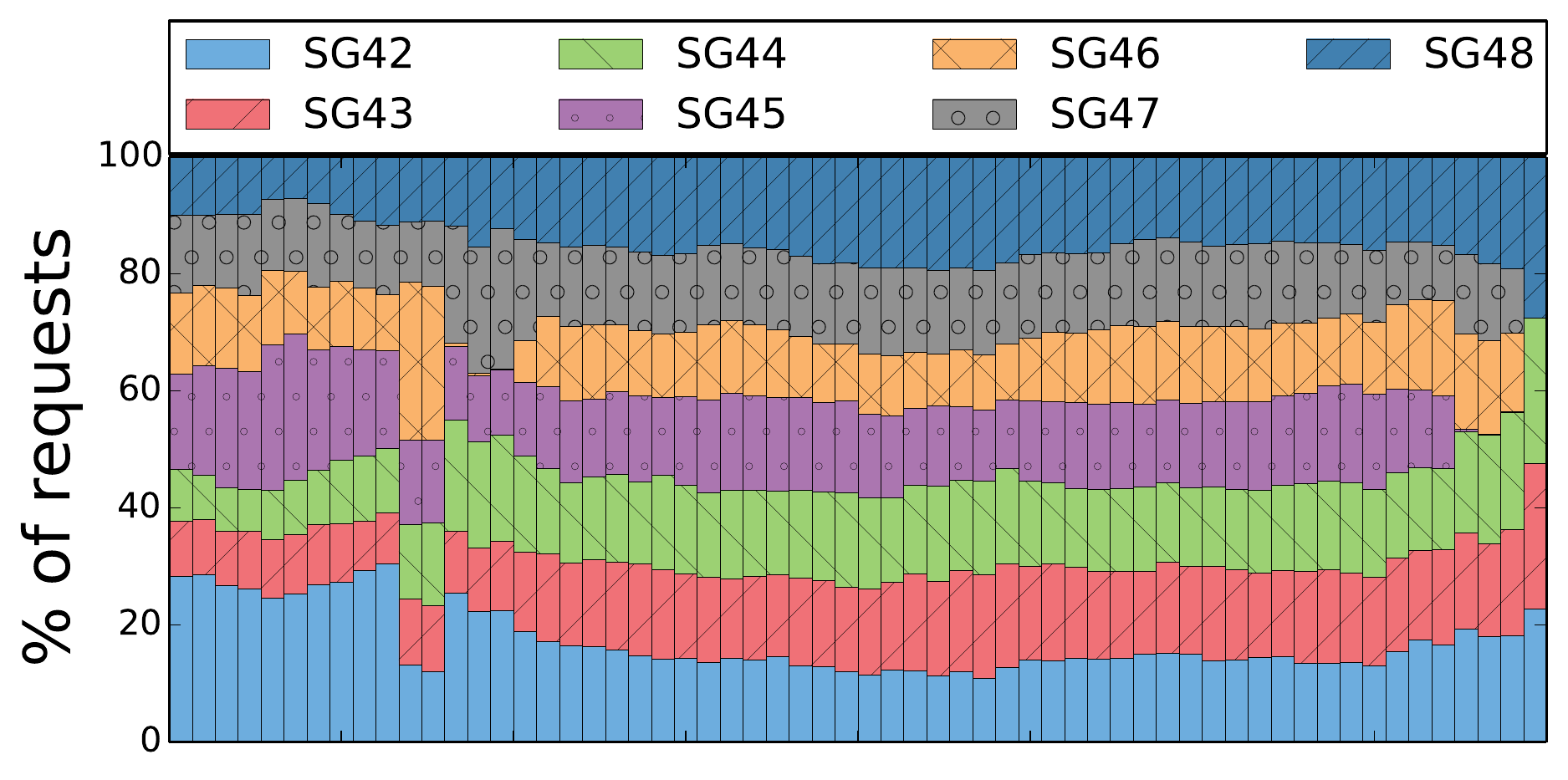}
 			\label{fig:tempSGsample-a}
        }
   \subfigure{
            \includegraphics[width=0.45\textwidth]{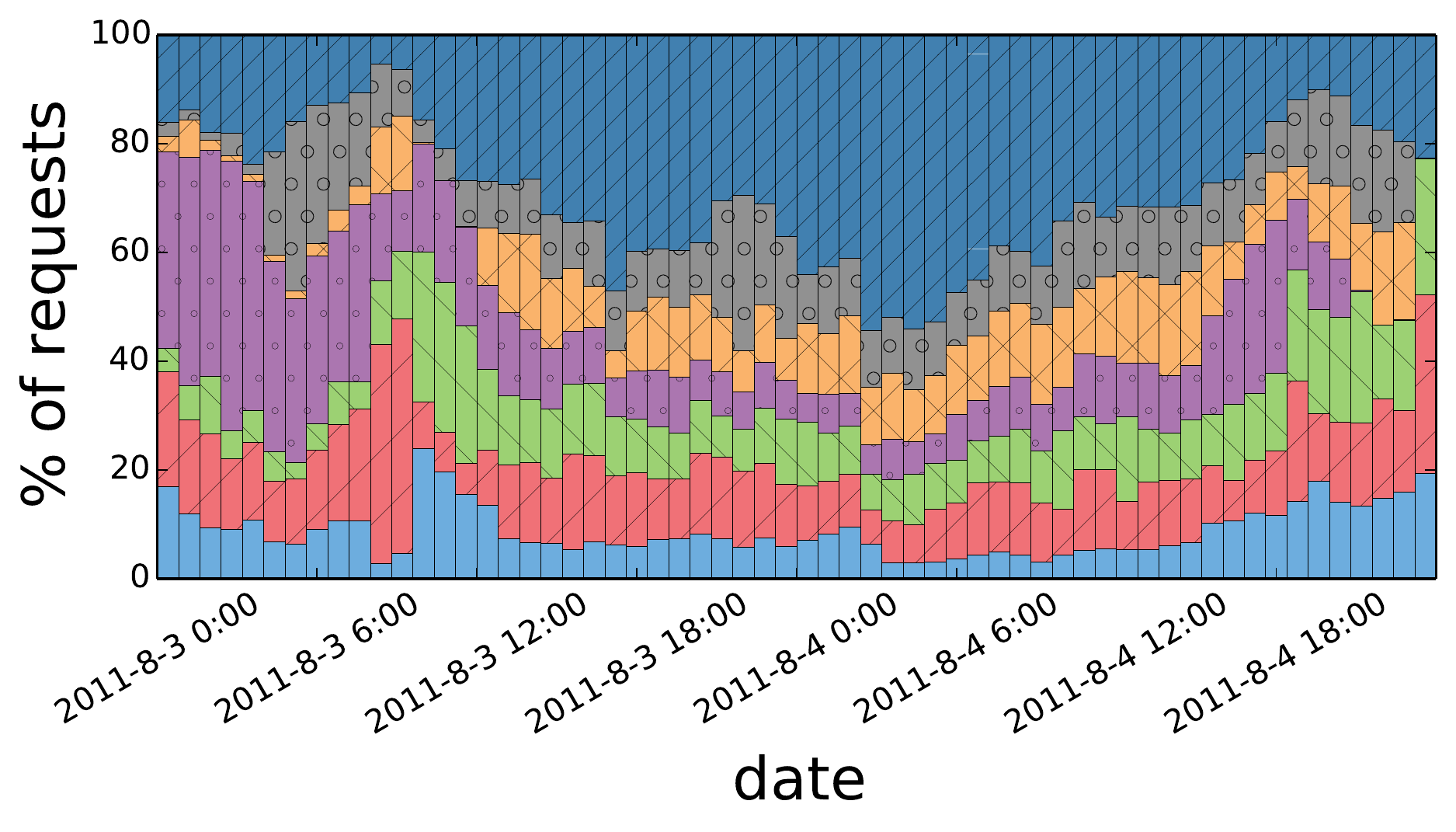}
 			\label{fig:tempSGsample-b}
        }
	\end{center}
	\vspace{-0.6cm}
\caption{The distribution of traffic load through each proxy  and censored traffic over time.}
	\label{fig:tempSG}
\end{figure}

We now compare the behavior of the 7 different proxies whose logs are included in our datasets. In Fig.~\ref{fig:tempSGsample-a}, we plot the traffic distribution across proxies, restricted to two days (August 3 and 4) to ease presentation. 
Note that the load is fairly distributed among the proxies, however, if one only considers
censored traffic (Fig.~\ref{fig:tempSGsample-b}), different behaviors become evident.
In particular, Proxy SG-48 is responsible for a
large proportion of the censored traffic, especially at certain times. 
One possible explanation is that different proxies follow different policies, or there could be a high proportion of  censored (or likely to be censored) traffic being redirected to proxy  SG-48 during one specific period of time.

We also consider the top-10 censored domain names in the period of time August 3 (12am)--August 4 (12am) and observe that the domain \url{metacafe.com} is \textit{always} censored and that almost all related requests (more than 95\%) are  processed {\em only} by proxy SG-48. This might be due to a domain-based traffic redirection process: in fact, we observed a very similar behavior for \url{skype.com} during the censorship peaks analysis presented earlier in Section~\ref{subsec:temporal}.

In order to verify our hypotheses, we evaluate the similarity between censored requests handled by each proxy. We do so by relying on the Cosine Similarity, defined as
{\small
$$cosine\_similarity(A,B)=\frac{\sum_{i=1}^{n}A_i\times B_i}{\sqrt{\sum_{i=1}^{n}(A_i)^2}\times \sqrt{\sum_{i=1}^{n}(B_i)^2}}$$
}

\noindent where $A_i$ and $B_i$ denote the number of requests for domain $i$ censored by proxies $A$ and $B$, respectively. Note that {\em cosine\_similarity} lies in the range [-1,1], with -1 indicating patterns that are not at all similar, and 1 indicating a perfect match. 
We report the values of {\em cosine\_simlarity} between the different proxies in Table~\ref{tab:domain_cosine_sg}. A few proxies exhibit high similarity, while others very low. This suggests that a few proxies are ``specialized'' in censoring specific types of content.

\begin{table}[ht!]
\centering
	\resizebox{0.99\columnwidth}{!}{
\begin{tabular}{|c|c|c|c|c|c|c|c|} \hline
~ & {\bf SG-42} & {\bf SG-43} & {\bf SG-44} & {\bf SG-45} & {\bf SG-46} & {\bf SG-47} & {\bf SG-48} \\ \hline
{\bf SG-42} & 1.0 & 0.5944 & 0.5424 & 0.3905 & 0.6134 & 0.2921 & 0.0896 \\ \hline
{\bf SG-43} & 0.5944 & 1.0 & 0.8226 & 0.4769 & 0.821 & 0.3138 & 0.0696 \\ \hline
{\bf SG-44} & 0.5424 & 0.8226 & 1.0 & 0.6177 & 0.8757 & 0.3003 & 0.0721 \\ \hline
{\bf SG-45} & 0.3905 & 0.4769 & 0.6177 & 1.0 & 0.4752 & 0.2316 & 0.6701 \\ \hline
{\bf SG-46} & 0.6134 & 0.821 & 0.8757 & 0.4752 & 1.0 & 0.3294 & 0.066 \\ \hline
{\bf SG-47} & 0.2921 & 0.3138 & 0.3003 & 0.2316 & 0.3294 & 1.0 & 0.0455 \\ \hline
{\bf SG-48} & 0.0896 & 0.0696 & 0.0721 & 0.6701 & 0.066 & 0.0455 & 1.0 \\ \hline
 \end{tabular} 
 }
\caption{Cross correlation of censored domains: Cosine similarity between different proxy servers (day: 2011-08-03).}
  \label{tab:domain_cosine_sg}
\end{table} 

We also look at the categories distribution of all requests across the different proxies and concentrate on two categories, ``Unavailable'' and ``None'', which show a peculiar distribution across the proxies (recall that categories have been discussed in Section~\ref{sec:general_characteristics}).
We note that the ``None'' category is \textit{only} observed on two different proxies (SG-43 and SG-48), while ``Unavailable'' is less frequently observed on these two. This suggests either different configuration of the proxies or a content specialization of the proxies.


\subsection{Denied vs. Redirected Traffic}
Requests are censored in one of two ways: they are either \emph{denied} or \emph{redirected}. If a request triggers a \policydenied exception, the requested page is not served to the client. Upon triggering \policyredirect, the request is redirected to another URL. For these requests, we only have information from the \emph{x-exception-id}   field (set to {\policyredirect}) and the \emph{s-action}  field  (set to {\em tcp\_policy\_redirect}). The \policyredirect exception is raised for a small number of hosts -- 11 in total. As reported in Table~\ref{tab:policy_redirect_hosts}, the most common URLs are \url{upload.youtube.com} and Facebook-owned domains. 

\begin{table}[t!]
\centering
\small
\begin{tabular}{|l|r|c|} \hline
 {\bf\em cs\_host} &    {\bf \# requests} & \% \\ \hline
upload.youtube.com   &   12,978 & 86.79\% \\ %
www.facebook.com     &   1,599 & 10.69\% \\  %
 ar-ar.facebook.com  &   264 & 1.77\% \\  %
 competition.mbc.net &   50 & 0.33\% \\ 
 sharek.aljazeera.net &  44 & 0.29\%  \\ 
 \hline
 \end{tabular} 
\caption{Top-5 hosts for \policyredirect requests in \df.}
\label{tab:policy_redirect_hosts}
\end{table}

Note that the redirection should trigger an additional request from the client to the redirected URL immediately after \policyredirect is raised. However, we found no trace of a secondary request coming right after (within a 2-second window). Thus, we conclude that the secondary URL is either hosted on a website that does not require to go through the filtering proxies (most likely, this site is hosted in Syria) or that the request is processed by proxies other than those in the dataset. 
Since the destination of the redirection remains unknown, we do not know whether or not redirections point to different pages, depending on the censored request.

\subsection{Category, String, and IP-based Filtering}
\label{subsec:cat_string_ip_censorship}
We now study the three main triggers of censorship decisions: URL categories, strings, and IP addresses.

\descr{Category-based Filtering.} According to Blue Coat's documentation~\cite{wikileaks}, proxies can associate a category to each request (in the {\em cs-categories} field), based on the corresponding URL, and this category can be used in the filtering rules.    
In the set of censored requests, we identify only two categories: a default category (named ``unavailable" or "none", depending on the proxy server), and a custom category (named ``Blocked sites; unavailable" or ``Blocked sites"). The custom category targets specific Facebook pages with a \policyredirect policy, accounts for 1,924 requests, and is discussed in detail in Section~\ref{sec:social_media}. All the other URLs (allowed or denied) are categorized to the default category, which is subject to a more general censorship policy, and captures the vast majority of the censored requests. The censored requests in the default category consist mostly of \policydenied  with a small portion (0.21\% either \texttt{PROXIED} or \texttt{DENIED}) of \policyredirect exceptions. We next investigate the policy applied within the default category.

\descr{String-based Filtering.} The filtering process is also based on particular strings included in the requested URL. 
In fact, the string-based filtering only relies on URL-related fields, specifically \emph{cs-host}, \emph{cs-uri-path} and  \emph{cs-uri-query}, which fully characterize the request. The proxies' filtering process is performed using a simple string-matching engine that detects any blacklisted substring in the URL.

We now aim to recover the list of strings that have been used to filter requests in our dataset. We expect that a string used for censorship should only be found in the set of censored requests and never in the set of allowed ones (for this purpose, we consider \texttt{PROXIED} requests separately from \texttt{OBSERVED} requests, since they do not necessarily indicate an allowed request, even when no exception is logged). 
In order to identify these strings, we use the following iterative approach: (1) Let $\mathcal{C}$ be the set of censored URLs and $\mathcal{A}$ the set of allowed URLs; (2) Manually identify a string $w$ appearing frequently in $\mathcal{C}$; (3) Let $N_\mathcal{C}$ and $N_\mathcal{A}$ be the number of occurrences of $w$ in $\mathcal{C}$ and $\mathcal{A}$, respectively; (4) If $N_\mathcal{C} >> 1$ and  $N_\mathcal{A} = 0$ then remove from $\mathcal{C}$ all requests containing $w$, add $w$ to the list of censored strings, and go to step 2.

To mitigate selection of strings that are unrelated to the censorship decision during the
manual string identification (step 2), we took a conservative approach by only considering non-ambiguous requests. For instance, we select simple requests, e.g., \texttt{HTTP GET new-syria.com/}, which only contains a domain name and has an empty path and an empty query field. Thus, we are sure that the string \url{new-syria.com} is the source of the censorship.

\descr{URL-based Filtering.}
Using the iterative process described above, we identify a list of 105 ``suspected'' domains, for which no request is allowed.  Table~\ref{tab:domain_blacklist} presents the top-10 domains in the list, according to the number of censored requests. We further categorized each domain in the list and show in Table~\ref{tab:categories_domain_blacklist} the top-10 categories according to the number of censored requests. Clearly, there is a heavy censorship of Instant Messaging software, as well as news, public forums, and user-contributed streaming media sites.

\begin{table}[t!]
\small
\centering
\begin{tabular}{|l|r|r|r|} \hline
{\bf Domain} & \multicolumn{1}{c|}{\bf  Censored} & \multicolumn{1}{c|} {\bf Allowed}& \multicolumn{1}{c|}{\bf Proxied}  \\ \hline 
metacafe.com & 1,278,583  (17.33\%) & 0  (0.00\%)  & 1,164  (0.03\%)   \\ 
skype.com & 503,932 (6.83\%)  & 0  (0.00\%) & 1,313  (0.04\%)   \\ 
wikimedia.org & 306,994  (4.16\%)  & 0  (0.00\%)  & 3,379  (0.10\%)   \\ 
.il &	112,369  (1.52\%) &	0  (0.00\%)  &	8,785  (0.25\%) \\ 
amazon.com & 62,759  (0.85\%)  & 0  (0.00\%)  & 345 (0.01\%)   \\ 
aawsat.com & 51,518  (0.70\%)  & 0 (0.00\%)  & 5,670  (0.16\%)   \\ 
jumblo.com & 23,214  (0.31\%)  & 0  (0.00\%)  & 0  (0.00\%)    \\ 
jeddahbikers.com & 21,274  (0.29\%)  & 0  (0.00\%)  & 130  (0.00\%)  \\ 
badoo.com & 14,502  (0.20\%) & 0   (0.00\%) & 358  (0.01\%)   \\ 
islamway.com & 14,408   (0.20\%)  & 0   (0.00\%) & 329  (0.01\%)   \\ 
\hline
\end{tabular} 
\caption{Top-10 domains suspected to be censored (number of requests and percentage for each class of traffic in \df).}
\label{tab:domain_blacklist}
\end{table}

\begin{table}[t!]
\small
\centering
\begin{tabular}{|l|rr|} \hline
{\bf Category (\#domains) } &   \multicolumn{2}{c|}{\bf Censored requests} \\ \hline
Instant Messaging (2)   & 47,116 & (16.63\%) \\    %
Streaming Media (6) & 39,282 & (13.87\%)  \\  %
Education/Reference (4) & 27,106 &  (9.57\%) \\ %
General News (62) & 8,700 & (3.07\%) \\ %
NA  (42) & 6,776 & (2.39\%) \\ %
Online Shopping  (2) &4,712 &  (1.66\%)  \\ %
Internet Services (6) & 2,964 & (1.05\%)  \\ %
Social Networking  (6) & 2,114 & (0.75\%)  \\ %
Entertainment (4) & 1,828 & (0.65\%)  \\ %
Forum/Bulletin Boards  (8) & 1,606 & (0.57\%)  \\ %
\hline
\end{tabular} 
\caption{Top-10 domain categories censored by URL (number of censored requests and percentage of censored traffic in \ds).}
\label{tab:categories_domain_blacklist}
\end{table}

\descr{Keyword-based Filtering.}
We also identify five keywords  that trigger censorship when found in the URL (\emph{cs-host}, \emph{cs-path} and \emph{cs-query} fields): {\em proxy}, {\em hotspotshield}, {\em ultrareach}, {\em israel}, and {\em ultrasurf}. We report the corresponding number of censored, allowed, and proxied requests in Table~\ref{tab:keyword_blacklist}. Four of them are related to anti-censorship technologies and one refers to Israel. 
Note that a large number of requests containing the keyword \emph{proxy} are actually related to seemingly ``non sensitive'' content, e.g., online ads content, tracking components or online APIs, but are nonetheless blocked. For instance, the Google toolbar API invokes a call to \texttt{/tbproxy/af/query}, which can be found on the \url{google.com} domain, and is unrelated to anti-censorship software. Nevertheless, this element accounts for 4.85\%  of the  \emph{censored} requests in the \ds dataset. Likewise, the keyword \emph{proxy} is also included in some online social networks' advertising components (see Section~\ref{sec:social_media}). 

\begin{table}[t!]
\small
\centering
\begin{tabular}{|l|r|r|r|} \hline
{\bf Keyword} & {\bf  Censored} & {\bf Allowed} & {\bf Proxied} \\ \hline
proxy & 3,954,795 (53.61\%) & 0 (0.00\%)  & 14,846 (0.42\%)  \\ 
hotspotshield &  126,127 (1.71\%) &   0 (0.00\%)& 816 (0.02\%) \\ 
ultrareach & 50,769 (0.69\%)  &  0 (0.00\%) & 68 (0.00\%)\\ 
israel & 48,119 (0.65\%) &  0 (0.00\%) &  477 (0.01\%) \\ 
ultrasurf & 31,483 (0.43\%)  &  0 (0.00\%) & 541 (0.02\%) \\ 
\hline
\end{tabular}
\caption{The list of 5 keywords identified as censored (fraction and number of requests for each class of traffic in \df). }
\label{tab:keyword_blacklist}
\end{table}

\descr{IP-based censorship.} We now focus on understanding whether some requests are censored based on IP address. To this end, we look at the requests for which the {\em cs-host} field is an IPv4 address  
and notice that some of the URLs of censored requests do not contain any meaningful information except for the IP address. As previously noted, censorship can be done at a country level, e.g., for Israel, as all .il domains are blocked. Thus, we consider the possibility of filtering traffic with destination in some specific geographical regions, based on the IP address of the destination host. 

We construct \dipv4, which includes the set of requests (from \df) for which the {\em cs-host} field is an IPv4 address. We geo-localize each IP address in \dipv4\ using the Maxmind GeoIP database.\footnote{\url{http://www.maxmind.com/en/country}} We then introduce, for each identified country, the corresponding censorship ratio, i.e., the number of censored requests over the total number of requests to this country.  Table~\ref{tab:ip_requests_countries} presents the censorship ratio for each country in \dipv4. Israel is by far the country with the highest censorship ratio, suggesting that it might be subject to an IP-based censorship.

\begin{table}[t!]
\small
\centering
\begin{tabular}{|l|c|r|r|} \hline 
{\bf Country} &  {\bf Censorship} & {\bf \# Censored } &  {\bf \# Allowed}   \\ 
 & {\bf Ratio (\%)} & & \\ \hline
Israel & 6.69 & 5,191 & 72,416 \\ 
Kuwait & 2.02 & 16 & 776 \\ 
Russian Federation & 0.64 & 959 & 149,161 \\ 
United Kingdom & 0.26 & 2,490 & 942,387 \\ 
Netherlands & 0.17 & 12,206 & 7,077,371 \\ 
Singapore & 0.13 & 19 & 14,768 \\ 
Bulgaria & 0.09 & 14 & 14,786 \\ 
\hline
\end{tabular} 
\caption{Censorship ratio for top censored countries in \dipv4.}
\label{tab:ip_requests_countries}
\end{table}

\begin{table}[t!]
\small
\centering
\begin{tabular}{|l|cc|cc|cc|} \hline
 &  \multicolumn{2}{c|}{{\bf Censored }}  & \multicolumn{2}{c|}{{\bf Allowed }} & \multicolumn{2}{c|}{{\bf Proxied }} \\ \hline
{\bf Subnet} &  {\bf \# req.} & {\bf \#  IPs} &  {\bf \# req.} & {\bf \#  IPs} &  {\bf \# req.} & {\bf \#  IPs}  \\ \hline
84.229.0.0/16 & 574 & 198 & 0 & 0 & 4 & 4 \\ \hline
46.120.0.0/15 & 571 & 11 & 5 & 1 & 0 & 0  \\ \hline
89.138.0.0/15 & 487 & 148 & 1 & 1 & 3 & 3  \\ \hline
212.235.64.0/19 & 474 & 5 & 325 & 1 & 0 & 0  \\ \hline
212.150.0.0/16 & 471 & 3 & 6,366 & 12 & 1 & 1   \\ \hline
\end{tabular} 
\caption{Top censored Israeli subnets.}
\label{tab:top_censored_israel_subnets}
\end{table}

Next, we focus on Israel and zoom in to the subnet level.\footnote{The list of IPv4 subnets corresponding to Israel is available from \url{http://www.ip2location.com/free/visitor-blocker}.} Table~\ref{tab:top_censored_israel_subnets} presents, for each of the top censored Israeli subnets, the number of requests and IP addresses that are censored and allowed. We identify two distinct groups: subnets that are almost always censored (except for a few exceptions of allowed requests), e.g., 84.229.0.0/16, and those that are either censored or allowed but for which the number of allowed requests is significantly larger than that of the censored ones, e.g. 212.150.0.0/16. 
One possible reason for a systematic subnet censorship could be related to blacklisted keywords. However, this is not the case in our analysis since the requested URL is often limited to  a single IP address (\emph{cs-uri-path} and \emph{cs-uri-query} fields are empty). We further check, using McAfee smart filter, that none but one (out of 1155 IP addresses) of 
the censored Israeli IP addresses are categorized as Anonymizer hosts. These results show then that IP filtering targets a few geographical areas, i.e., Israeli subnets.  

\subsection{Summary}
The analysis presented in this section has shown evidence of domain-based traffic redirection between proxies. A few proxies seem to be specialized in censoring specific domains and type of content.  Also, our findings suggest that the censorship activity reaches peaks mainly because of unusually high demand for Instant Messaging Software websites (e.g., Skype), which are blocked in Syria. 
We found that censorship is based on four main criteria: URL-based filtering, keyword-based filtering, destination IP address, and a custom category-based censorship (further discussed in the next section). 
The list of blocked keywords and domains demonstrates the intent of Syrian censors to block political and news content, video sharing, and proxy-based censorship-circumvention technologies. Finally, Israeli-related content is heavily censored as the keyword \emph{Israel}, the \url{.il} domain, and some Israeli subnets are blocked.

\section{Censorship of Social Media}
\label{sec:social_media}

In this section, we analyze the filtering and censorship of Online Social Networks (OSNs) in Syria. Social media have often been targeted by censors, e.g., during the recent uprisings in the Middle East and North Africa. In Syria, according to our logs, popular OSNs like Facebook and Twitter are not \textit{entirely} censored and most traffic is allowed. However, we observe that a few specific keywords (e.g., \emph{proxy}) and a few pages (e.g., the `Syrian Revolution' Facebook page) are blocked, thus suggesting a targeted censorship.

We select a representative set of social networks containing the top 25 social networks according to \url{alexa.com} as of November 2013, and add 3 social networks popular in Arabic-speaking countries: \url{netlog.com}, \url{salamworld.com}, and \url{muslimup.com}. For each of these, we extract the number of allowed, censored and proxied requests in \df, and report the top-10 censored social networks 
in Table~\ref{tab:social_network_summary}. 

\begin{table}[t!]
\centering
\small
\begin{tabular}{|l|r|r|r|} \hline
    {\bf OSN} &  \multicolumn{1}{c|} {\bf Censored} & \multicolumn{1}{c|}{\bf Allowed} & \multicolumn{1}{c|}{\bf Proxied} \\ \hline
facebook.com & 1,616,174  (21.91\%)  & 17.70M  (2.53\%) & 125,988  (3.60\%)    \\ 
badoo.com & 14,502  (0.20\%)  & 0  (0.00\%) & 358  (0.01\%)  \\ 
netlog.com & 9,252  (0.13\%)  & 0  (0.00\%) & 2,227  (0.06\%)    \\ 
linkedin.com & 7,194  (0.10\%)  & 186,047   (0.03\%) & 1,723  (0.05\%)   \\ 
skyrock.com & 3,307  (0.04\%)  & 7,564  (0.00\%)  &  11  (0.00\%)   \\ 
hi5.com & 2,995  (0.04\%)  & 210,411  (0.03\%)  & 463  (0.01\%)    \\ 
twitter.com & 163  (0.00\%)  &  2.83M  (0.40\%) & 14,654  (0.42\%)    \\ 
ning.com &  6  (0.00\%) & 41,993  (0.01\%)  & 69  (0.00\%)   \\ 
meetup.com & 3  (0.00\%) &  108  (0.00\%) & 0   (0.00\%)   \\ 
flickr.com & 2  (0.00\%)  & 383,212  (0.05\%)  & 3179  (0.09\%)    \\ 
\hline
 \end{tabular} 
\vspace{-0.2cm}
\caption{Top-10 censored social networks in \df (number and percentage of requests for each class of traffic).}
\vspace{-0.4cm}
\label{tab:social_network_summary}
\end{table} 

We find no evidence of systematic censorship for most sites (including last.fm, MySpace, Google+, Instagram, and Tumblr), as all requests are allowed. However, for a few social networks (including Facebook, Linkedin, Twitter, and Flickr) many requests are blocked. Several requests are censored based on blacklisted keywords (e.g., \emph{proxy, Israel}), thus suggesting that the destination domain is not the actual reason of censorship. However, requests to Netlog and Badoo are never allowed and there is only a minority of requests containing blacklisted keywords, which suggests that these domains are always censored. In fact, both \url{netlog.com} and \url{badoo.com} were identified in the list of domains suspected for URL-based filtering, described in Section~\ref{subsec:cat_string_ip_censorship}.

\descr{Facebook.} The majority of requests to Facebook are allowed, yet \url{facebook.com} is one of the most censored domains. As we explain below, censored requests can be classified into two groups: (i) requests to Facebook pages with sensitive (political) content, and (ii) requests to the social platform with the blacklisted keyword {\em proxy}.

\descr{Censored Facebook pages.} Several Facebook pages are censored for political reasons and are identified by the proxies using the custom category
``Blocked Sites.'' Requests to those pages trigger a {\em policy\_redirect} exception, thus redirecting the user to a page unknown to us. Interestingly, Reporters Without Borders \cite{RWB} stated that
``the government's cyber-army, which tracks dissidents on online social networks, seems to have stepped up its activities since June 2011. Web pages that support the demonstrations were flooded with pro-Assad messages.''  While we cannot infer the destination of redirection, we argue that this mechanism could technically serve as a way to show specific content addressing users who access targeted Facebook pages.

Table~\ref{tab:Filtered_facebook_pages} lists the Facebook pages we identify in the logs that fall into the custom category. All the requests identified as belonging to the custom category are censored.  However, we find that not all requests to the \url{facebook.com/<censored_page>} pages are correctly categorized as ``Blocked Site.'' For instance, \url{www.facebook.com/Syrian.Revolution?ref=ts} is, but \url{www.facebook.com/Syrian.Revolution?ref=ts} is, but \url{www.facebook.com/Syrian.Revolution?ref=ts&__a=11&ajaxpipe=1&quickling[version]=414343%3B0}
is not, thus suggesting that the categorization rules targeted a very narrow range of specific \emph{cs-uri-path} and \emph{cs-uri-query} combinations. As shown in Table~\ref{tab:Filtered_facebook_pages}, many requests to targeted Facebook pages are allowed and no allowed request is categorized as ``Blocked Site.'' We also identify successful requests sent to Facebook pages such as Syrian.Revolution.Army, Syrian.Revolution.Assad, Syrian.Revolution.Caricature and ShaamNewsNetwork, which are 
not categorized as ``Blocked Site'' and are allowed. Finally, proxied requests are sometimes, but not always, categorized as ``Blocked Site'' (e.g., all the requests for the Syrian.revolution page).

\begin{table}[t!]
\small
\centering
\begin{tabular}{|l|r|r|r|} \hline
{\bf  Facebook page} &  {\bf \# Censored} & {\bf \# Allowed} &  {\bf \# Proxied} \\ \hline
 Syrian.Revolution & 1461 & 891 & 16   \\ 
 Syrian.revolution & 0 & 0 & 25   \\ 
 syria.news.F.N.N & 191 & 165 & 1   \\ 
 ShaamNews & 114 & 3944 & 7   \\ 
 fffm14 & 42 & 18 & 0   \\ 
 barada.channel & 25 & 9 & 0   \\ 
 DaysOfRage & 19 & 2 & 0   \\ 
 Syrian.R.V & 10 & 6 & 0   \\ 
 YouthFreeSyria & 6 & 0 & 0   \\ 
 sooryoon & 3  & 0 & 0   \\ 
 Freedom.Of.Syria & 3 & 0 & 0 \\
 SyrianDayOfRage & 1 & 0 & 0 \\
\hline
\end{tabular}
\caption{Top blocked Facebook pages in \df.}
\label{tab:Filtered_facebook_pages}
\end{table}

\descr{Social plugins.} Facebook provides so-called social plugins (one common example is the {\em Like} button), which is loaded by web pages to enable interaction with the social platform. Some of the URLs used by these social plugins include  the keyword {\em proxy} in the  \emph{cs-uri-path} field 
or in the  \emph{cs-uri-query} field, 
and this automatically raises the  \policydenied exception whenever the page is loaded.

Table~\ref{tab:facebook_plugins} reports, for each of the top-10 social plugin elements, the fraction of the Facebook traffic and the number of requests for each class of traffic. The top two censored social plugin elements (\emph{/plugins/like.php} and \emph{/extern/login\_status.php}) account for more than 80\% of the censored traffic on the \url{facebook.com} domain, while the 10 social plugin elements we consider account for 99.9\% (1,612,835) of the censored requests on the \url{facebook.com} domain.
To conclude, the large number of censored requests on the \url{facebook.com} domain is in fact mainly caused by social plugins elements that are not related with censorship circumvention tools or any political content.

\begin{table}[t!]
\centering
\resizebox{1.0\columnwidth}{!}{
\begin{tabular}{|l|r|r|r|} \hline
    {\bf Social plug-in} &  \multicolumn{1}{c|} {\bf  Censored} & \multicolumn{1}{c|}{\bf  Allowed} & \multicolumn{1}{c|}{\bf  Proxied} \\ \hline
/plugins/like.php & 694,788  (43.04\%)  & 0 (0.00\%) & 8,919  (7.08\%) \\ 
/extern/login\_status.php & 629,495  (38.99\%)  & 0 (0.00\%) & 3,502  (2.78\%) \\
/plugins/likebox.php & 77,244  (4.78\%) & 0 (0.00\%) & 3555  (2.82\%) \\ 
/plugins/send.php & 70,146  (4.35\%)  & 0 (0.00\%) & 272  (0.22\%) \\
/plugins/comments.php & 54,265  (3.36\%)   & 0 (0.00\%) & 331  (0.26\%) \\
/fbml/fbjs\_ajax\_proxy.php & 42,649  (2.64\%)  & 0 (0.00\%) & 43  (0.03\%) \\ 
/connect/canvas\_proxy.php & 40,516  (2.51\%)   & 0 (0.00\%) & 37  (0.03\%) \\ 
/ajax/proxy.php & 1,544  (0.10\%)  & 0 (0.00\%) & 6  (0.00\%) \\
/platform/page\_proxy.php & 1,519  (0.09\%)  & 0 (0.00\%) & 4  (0.00\%) \\
/plugins/facepile.php & 669  (0.04\%)  & 0 (0.00\%) & 4  (0.00\%) \\
\hline
\end{tabular}
}
\caption{Top-10 Facebook social plugin elements in \df (fraction of Facebook  traffic and number of requests).}
\label{tab:facebook_plugins}
\end{table}

\descr{Summary.} We have studied the censorship of 28 major online social networks
and found that most of them are not censored, unless requests
contain blacklisted keywords (such as {\em proxy}) in the URL. This is particularly
evident looking at the large amount of Facebook requests that are censored 
due to the presence of {\em proxy} in the query. Using a custom category, the censors also target a selected number of Facebook pages, without blocking all traffic to the site, thus making censorship and surveillance harder to detect (as independently reported in~\cite{oni}).

\section{Anti-censorship Technologies}\label{sec:anti}
We now investigate the usage (and effectiveness) of censorship-circumvention technologies  based on our dataset.

\subsection{Tor}\label{sec:tor}
According to the logs, access to the Tor project website and the majority of Tor traffic were allowed in July/August 2011. In fact, access to the Tor network was first reported to be blocked on December 16, 2012~\cite{tor-syria}.

Tor traffic can be classified into two main classes: (1) HTTP signaling, e.g., establishing connections with Tor directories, which we denote as $Tor_{http}$, and (2) establishing Tor circuits and data transfer, denoted as $Tor_ {onion}$. 
To identify Tor traffic, we extract Tor relays' IP addresses and port numbers from the Tor server descriptors and network status files (available from \url{https://metrics.torproject.org/formats.html}). We then match the extracted $<$node IP, port, date$>$ triplets to the requests in \df\ to identify Tor traffic. We further isolate HTTP signaling messages by identifying all HTTP requests to Tor directories, e.g.,  \texttt{/tor/server/authority.z} or  \texttt{/tor/keys}.\footnote{See \url{https://gitweb.torproject.org/torspec.git?a=blob_plain;hb=HEAD;f=dir-spec-v2.txt} for a full description.}
This does not take into account the connections via Tor bridges: there is no public list of them (bridges are used to overcome filtering of connections to known Tor relays), however,  Tor relays were not filtered in Syria as of 2011, thus users did not actually need to use bridges.

\begin{figure}[t!]
  \begin{center}
  \subfigure[]{
            \includegraphics[width=0.38\textwidth]{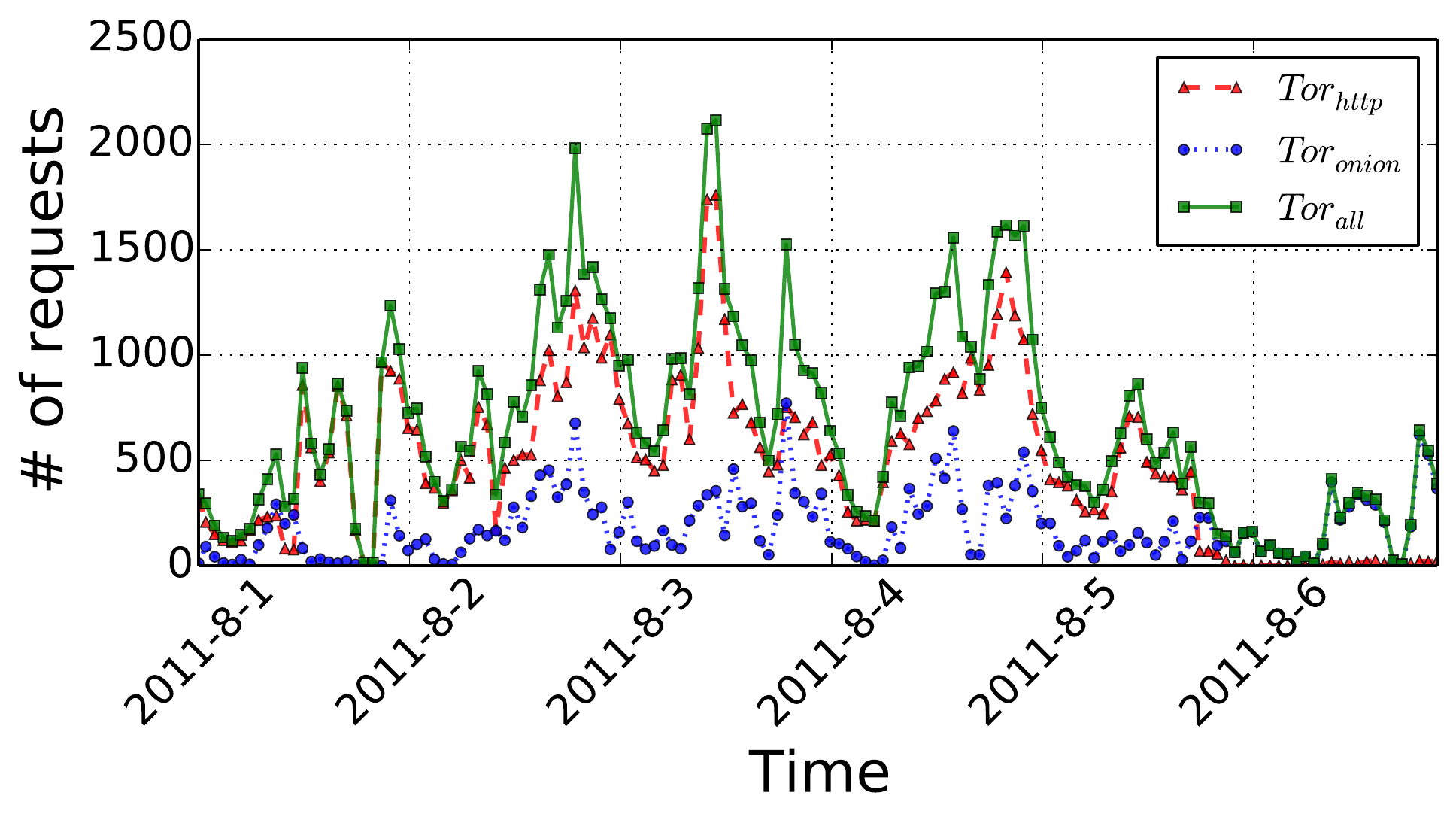} %
 			\label{fig:tor_traffic}
        }
   \subfigure[]{
            \includegraphics[width=0.38\textwidth]{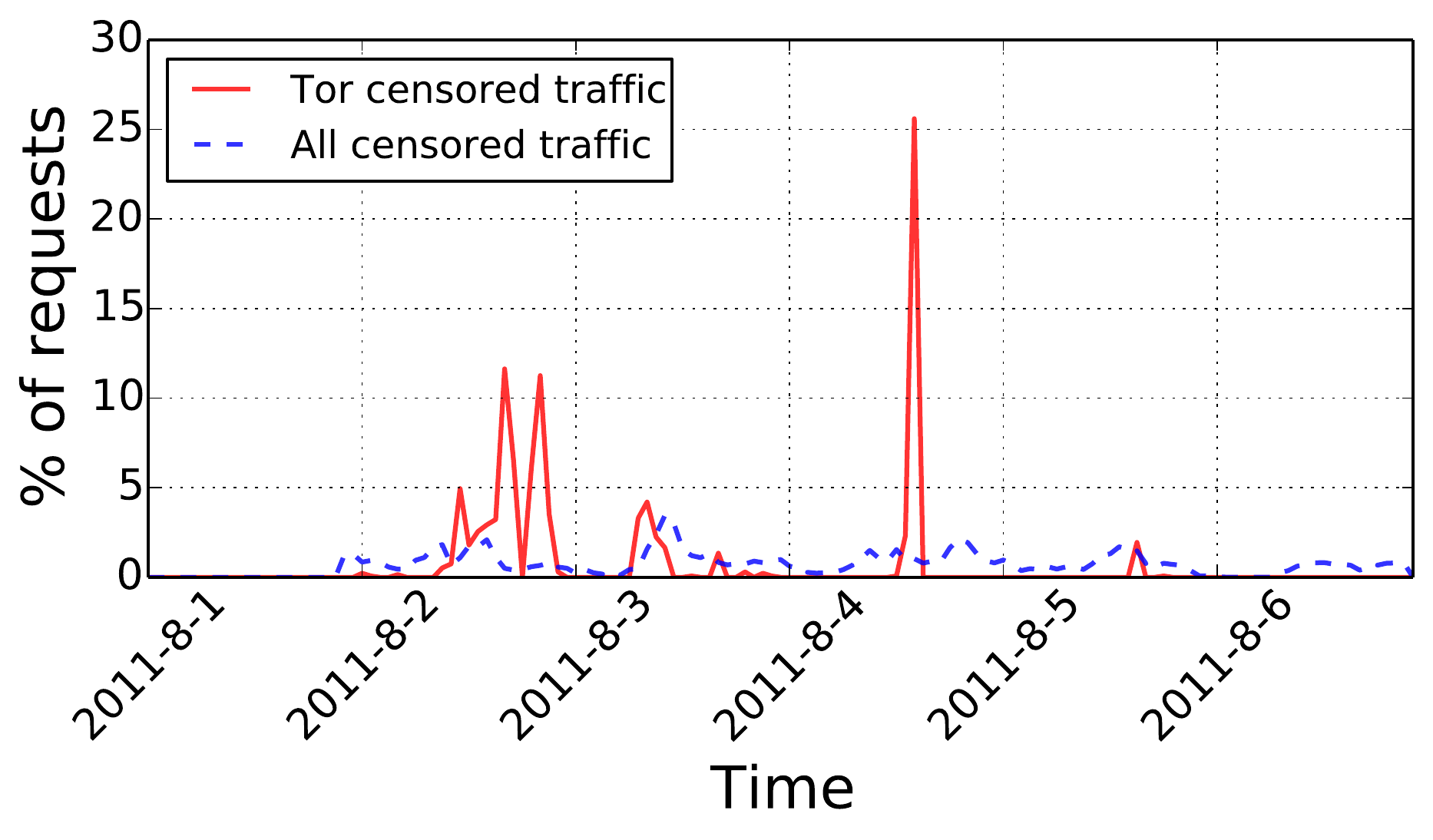}
 			\label{fig:tor_sg44censored}
        }
        \vspace*{-0.2cm}
        \caption{(a) Number of Tor related requests per hour from August 1-6 in \df.; (b) Percentage of all censored traffic and Tor censored traffic by Proxy SG-44.}
	\end{center}

\end{figure}

We identify 95K requests to 1,111 different Tor relays, 73\% of which belong to $Tor_{http}$. Only a small fraction (1.38\%) of the requests are censored and 16.2\% of them generate TCP errors. Figure~\ref{fig:tor_traffic} shows the number of requests for August 1-6. The traffic has several peaks, in particular on August 3, when several protests were taking place.
We observe that 99.9\% of censored Tor traffic is blocked by a single proxy (SG-44), even though the overall traffic is uniformly distributed across all the 7 proxies. The other 0.01\% of the traffic is censored by SG-48. This finding echoes our earlier discussion on the specialization of some proxies. 

We also analyze the temporal pattern of Tor censored traffic and compare it to the overall censored requests of SG-44. As shown in 
Figure~\ref{fig:tor_sg44censored}, Tor censoring exhibits a higher variance.
While censoring $Tor_{http}$ is technically simple, as it only requires matching of regular expressions against the HTTP requests, identifying and censoring $Tor_{onion}$ is more challenging, as it involves encrypted traffic. However, only $Tor_{onion}$ traffic is censored according to the logs, while $Tor_{http}$ is always allowed.

\begin{figure}[t!]
 \begin{center}
 \includegraphics[width=0.35\textwidth]{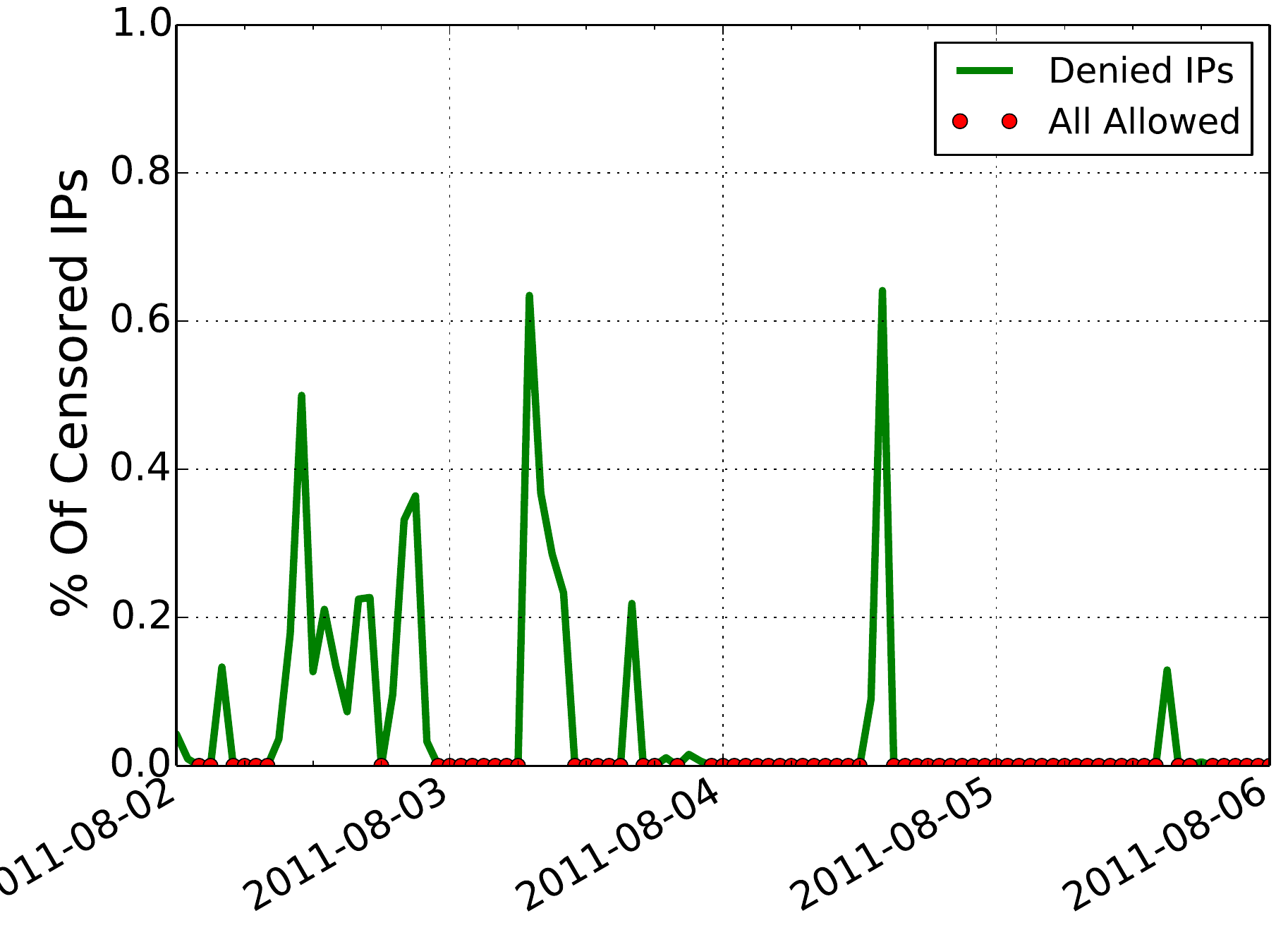}
\caption{Ratio of (re)censored IPs SG-44.}
\label{fig:tor_sg44_spec}
 \vspace{-0.2cm}
\end{center}
\end{figure}

Next, we verify whether the Tor censorship is consistent (i.e., whether the blocked Tor relay is re-allowed later). To do so, we proceed as follows: 
First, we create a set of all censored Tor relay addresses denoted \texttt{Censored-IPs}.  
Then, for each one hour time window considered as a time bin $k$, we create the set of all Tor relays IP addresses that were allowed, we denote \texttt{Allowed-IPs (k)}. For each time bin $k$, we calculate the relative overlap between the censored Tor relay nodes and the Tor relays allowed at bin $k$ as: 
\[\small R_{filter}(k)= 1 - \frac{|\texttt{Censored-IPs}~ \cap ~\texttt{Allowed-IPs(k)} |}{|\texttt{Censored-IPs}|} \vspace{-0.1cm}\]

\noindent where $|\cdot|$ denotes the set size. Figure~\ref{fig:tor_sg44_spec} depicts the variation of $R_{filter}$ as a function of time. Note that $R_{filter}(k)$ equals 0 if, at a specific time bin $k$, all connections to \texttt{Censored-IPs} are allowed (line curve) or none of the IPs in Allowed are in Censored (circle curve). $R_{filter}(k)$ equals 1 if none of the connections to  \texttt{Censored-IPs} have been allowed at a specific time bin $k$. 

The high variance of $R_{filter}$ shows that the censorship is inconsistent. In fact, the beginning of this plot is either dominated by red circles (all traffic is allowed) or few small peaks showing a very mild censorship. Then few successive peaks with a high variance of blockage appear for several hours, followed by a lull at the night of 03 of August. The same scenario is repeated several times, alternating between aggressive censorship and more mild period. This behavior  is hard to explain as some IPs are alternating between blocked and allowed status. One explanation might be a testing phase that is carried in a single router and for a short period of time to test a new censorship approach (in this scenario the newly deployed Tor censorship).  Or, this censorship is based on fields that are not logged by the appliance and hence inaccessible to us.

\subsection{Web Proxies and VPNs}
As already mentioned, access to web/socks proxies is censored, as demonstrated by an aggressive filtering of requests including the keyword {\em proxy}.
In the following, we use ``proxies'' to refer to the Blue Coat appliances whose logs we study in this paper, and ``web proxies'' to refer to services used to circumvent censorship.

To use web proxies, end-users need to configure their browsers or their network interfaces, or rely on tools (e.g., Ultrasurf) that automatically redirect all HTTP traffic to the web proxy. Some web proxies support encryption, and create a SSL-based encrypted HTTP tunnel between the user and the web proxy. Similarly, VPN tools (e.g., Hotspot Shield) are often used to circumvent censorship, again, by relaying traffic through a VPN server. 

We analyze the usage of VPN and web proxy tools and find that 
a few of them are very popular among Internet users in Syria. However, as discussed in Section \ref{subsec:cat_string_ip_censorship}, keywords like {\em ultrasurf} and {\em hotspotshield} are heavily monitored and censored.
Nonetheless, some web proxies and VPN software (such as Freegate, GTunnel and GPass) do not include the keyword {\em proxy} in request URLs  and are therefore not censored. Similarly, we do not observe any censorship activity triggered by the keyword {\em VPN}. 

\begin{figure}[t!]
  \begin{center}
  \subfigure[]{
            \includegraphics[width=0.35\textwidth]{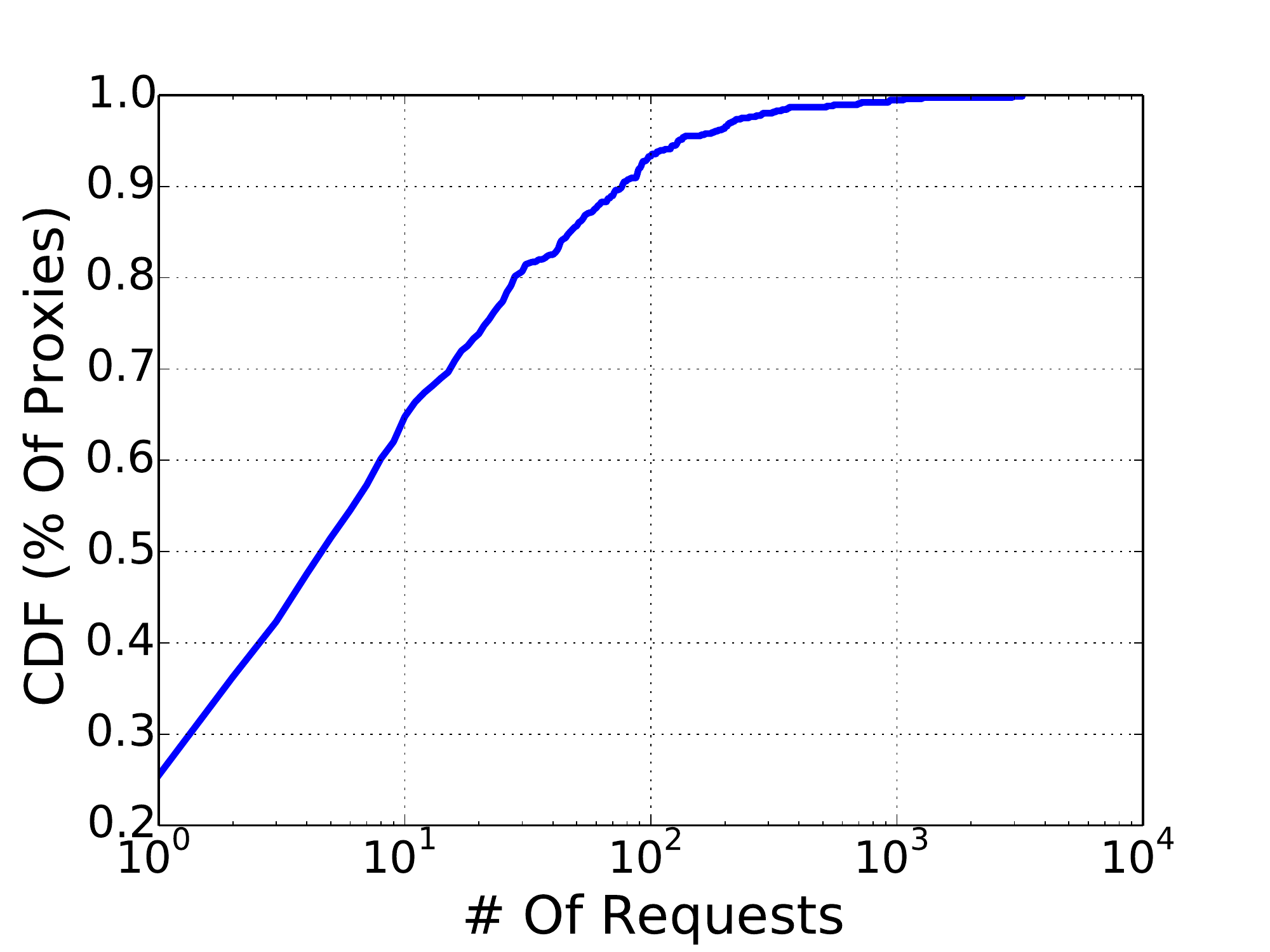}
 			\label{fig:cdf_anonymizers}
                }
           \subfigure[]{
            \includegraphics[width=0.35\textwidth]{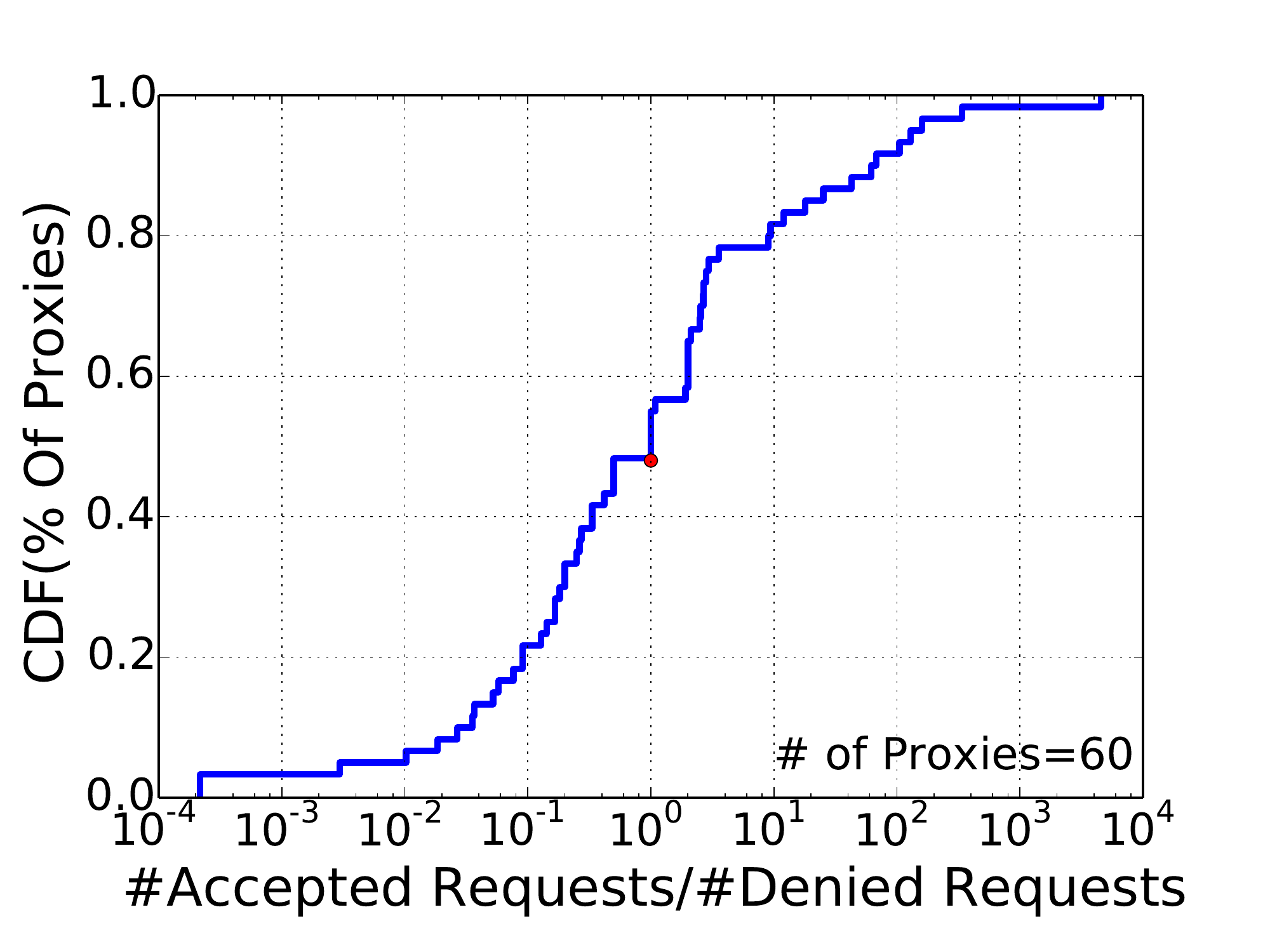}
 			\label{fig:proxy_inconsistency}
        }
        \vspace*{-0.2cm}
        \caption{(a) CDF of the number of requests for identified ``anonymizer'' hosts; (b) Ratio of Allowed versus Censored number of requests for identified ``anonymizer'' hosts.}
	\end{center}
\vspace*{-0.2cm}
\end{figure}

Next, we focus on the domains that are categorized as ``Anonymizers'' by McAfee's TrustedSource tool,  including both web proxies and VPN-related hosts.
In the \ds\ dataset, there are 821 ``Anonymizer'' domains, which are the target of 122K requests (representing 0.4\% of the total number of requests). 92.7\% of these hosts (accounting for 25\% of the requests) are never filtered. Figure~\ref{fig:cdf_anonymizers} shows the CDF of the number of requests sent to each of those allowed hosts. Less than 10\% of these hosts receive more than 100 requests, suggesting that only a few popular services attract a high number of the requests.

Finally, we look at the 7.3\% of the identified ``Anonymizer'' hosts, for which some of the requests are censored. We calculate the ratio between the number of allowed requests in \df and the number of censored requests in \dd. Figure~\ref{fig:proxy_inconsistency} shows the CDF of this censorship ratio. We observe a  non-consistent policy for whether a request is allowed or censored, with more than 50\% of the proxies showing a higher number of allowed requests than censored requests. This suggests that these requests are not censored based on their IP or hostname, but rather on other criteria, e.g., the inclusion of a blacklisted keyword in the request.

In conclusion, while some services (such as Hotspot Shield) are  heavily censored, other, less known, services are not, unless related requests contain blacklisted keywords. This somehow introduces a trade-off between 
the ability to bypass censorship and promoting censorship- and surveillance-evading tools (e.g., in web searches) by including keywords such as {\em proxy} in the URL.

\subsection{Peer-to-Peer networks}

The distributed architecture of peer-to-peer networks makes them, by nature, more resilient to censorship: users obtain content from peers and not from a server, which makes it harder to locate and block content. Shared data is usually identified by a unique identifier (e.g., info hash in BitTorrent), and these identifiers are useless to censors unless mapped back to, e.g., the description of the corresponding files. However, resolving these identifiers is not trivial: content can be created by anyone at any time, and the real description can be distributed in many different ways, publicly and privately. 

To investigate the use of peer-to-peer networks as a way to access censored content, we look for signs of BitTorrent traffic. We find a total of 338,168 BitTorrent announce requests from 38,575 users (\red{BitTorrent uses a 20-byte peer\_id field to identify peers, which we use to count unique users}) for 35,331 unique contents in the \df\ dataset.\footnote{BitTorrent clients send announce requests to BitTorrent servers (aka trackers) to retrieve a list of IP addresses from which the requested content can be downloaded.} Most of these requests (99.97\%) are allowed. Censored requests can be explained with the occurrence of blacklisted keywords, e.g., {\em proxy}, in the request URL. For instance, all announce requests sent to the tracker on \emph{tracker-proxy.furk.net} are censored. 

Using the hashes of torrent files provided in the announce messages, we crawl  \url{torrentz.eu} and \url{torrentproject.com} to extract the titles of these torrent files, achieving a success rate of 77.4\%. 
The five blacklisted keywords, reported in Table \ref{tab:keyword_blacklist}, are actually present in the titles of some of the BitTorrent files, yet the associated announce requests are allowed. We do not find any content that can be directly associated with sensitive topics like ``Syrian revolution'' or ``Arab spring'' (these files may still be shared via BitTorrent without publicly announcing the content), however, we identify content relating to anti-censorship software, such as UltraSurf (2,703 requests for all versions), HideMyAss (176 requests), Auto Hide IP (532 requests) and anonymous browsers (393 requests). Our findings suggest that peer-to-peer networks are indeed used by users inside Syria to circumvent censorship to a certain extent. %
Also note that BitTorrent is used to download Instant Messaging software, such as Skype, MSN messenger, and Yahoo! Messenger, which cannot be downloaded directly from the official download pages due to censorship.

\subsection{Google Cache}

When searching for terms in Google's search engine, the result pages allow access to  cached versions of suggested pages.  While Google Cache is not intended as an anti-censorship tool, a simple analysis of the logs shows that it provides a way to access content that is otherwise censored. 

We identify a total of 4,860 requests accessing Google's cache on  \url{webcache.googleusercontent.com} in the \df dataset. Only 12 of them are censored due to an occurrence of a blacklisted keyword in the URL, and a single request to retrieve a cached version of \url{http://ar-ar.facebook.com/SYRIANREVOLUTION.K.N.N} has \policydenied, although it is not categorized as a ``Blocked Site.'' However, the rest of the requests are allowed. Interestingly, some of the allowed requests, although small in number, relate to cached versions of webpages that are otherwise censored, such as \url{www.panet.co.il}, \url{aawsat.com}, \url{www.facebook.com/Syrian.Revolution}, and \url{www.free-syria.com}. 
While the use of Google cache to access censored content is obviously limited in scope, the logs actually suggest that it is very effective. Thus, when properly secured with HTTPS, Google cache could actually serve as a way to access censored content.

\subsection{Summary}
The logs highlighted that Syrian users do resort to censorship circumvention tools, with a relatively high effectiveness. While some tools and websites are monitored and blocked (e.g., Hotspot Shield), many others are successful in bypassing censorship. Our study also shows that 
some tools that were not necessarily designed as circumvention tools, such as BitTorrent and Google cache, could provide additional ways to access censored content if proper precautions are taken, especially considering that Syrian ISPs started to block Tor relays and bridges in December 2012.

\section{Discussion}
\label{sec:discussion}

\descr{Economics of Censorship.} Our analysis shows that Syrian authorities deploy several techniques to filter Internet traffic, ranging from blocking entire subnets to filtering based on specific keywords. This range of techniques can be explained by the cost/benefit tradeoff of censorship, as described, e.g., by Danezis and 
Anderson~\cite{danezis}. While censoring the vast majority of the Israeli network -- regardless of the actual content -- can be explained on geo-political grounds, completely denying the access to social networks, such as Facebook, could generate unrest. For instance, facing the ``Arab spring'' uprisings, the Syrian authorities decided to allow access to Facebook, Twitter, and Youtube in February 2011.
Nonetheless, these websites are monitored and selectively censored. Censors might aim at a more subtle control of the Internet, by only denying access to a predefined set of websites, as well as a set of keywords. This shift is achievable as the proxy appliances seamlessly support Deep Packet Inspection (DPI), thus allowing fine-grained censorship in real-time.    

\descr{Censorship's target.} Censored traffic encompasses a large variety of content, mostly aiming to prevent users from using Instant Messaging software (e.g., Skype), video sharing websites (e.g., \url{metacafe.com},  \url{upload.youtube.com}), Wikipedia, as well as sites related to news and opposition parties (e.g., \url{islammemo.cc}, \url{alquds.co.uk}). Censors also deliberately block any requests related to a set of predefined anti-censorship tools (e.g., `proxy'). This mechanism, however, has several side effects as it denies the access to any page containing these keywords, including those that have nothing to do with censorship circumvention.

\descr{Censorship Circumvention.} We also highlighted that users attempt to circumvent censorship. One interesting way is using BitTorrent to download anti-censorship tools such as UltraSurf as well as Instant Messaging software. Users also rely on known censorship-evading technologies, such as web/socks proxies and Tor.

\section{Conclusion}\label{sec:conclusion}

This paper presented a measurement analysis of Internet filtering in Syria.
We analyzed 600GB worth of logs produced by 7 Blue Coat SG-9000 proxies in Summer 2011 and, 
by extracting  information about processed requests for both censored and allowed traffic, we provided a detailed, first-of-a-kind snapshot of a real-world censorship ecosystem. We uncovered the presence of a relatively stealthy yet quite targeted filtering, which relies on IP addresses to block access to entire subnets, on domains to block specific websites, and on keywords to target specific content. 
Keyword-based censorship (e.g., denying all requests containing the word `proxy') also produces  collateral damage, as many requests are blocked even if they do not relate to sensitive content.
Finally, we showed that Instant Messaging software is heavily censored, that filtering of social media appears to be limited to specific pages, and that Syrian users try to circumvent censorship using web/socks proxies, Tor, VPNs, and BitTorrent.

\balance	

%
\end{document}